\documentclass[11pt]{article}

\usepackage[total={6.5in,9in},top=1in, left=1in]{geometry}
\usepackage{psfrag}
\usepackage{setspace}
\usepackage{graphics}
\usepackage{subfigure}
\usepackage{verbatim}
\usepackage{amssymb}
\usepackage{amsmath}
\usepackage{lscape}
\usepackage{epsfig}
\usepackage{color}
\usepackage{jneurosci}
\singlespace
\newcommand{\beq}{\begin{equation}}
\newcommand{\eeq}{\end{equation}}
\newcommand{\beqr}{\begin{eqnarray}}
\newcommand{\eeqr}{\end{eqnarray}}
\newcommand{\beqrn}{\begin{eqnarray*}}
\newcommand{\eeqrn}{\end{eqnarray*}}
\newcommand{\beqn}{\begin{equation*}}
\newcommand{\eeqn}{\end{equation*}}
\newcommand{\bei}{\begin{itemize}}
\newcommand{\beii}{\begin{itemize} \item}
\newcommand{\eei}{\end{itemize}}
\newcommand{\ben}{\begin{enumerate}}
\newcommand{\een}{\end{enumerate}}
\newcommand{\bes}{\begin{small}}
\newcommand{\ees}{\end{small}}
\newcommand{\bec}{\begin{center}}
\newcommand{\eec}{\end{center}}

\newcommand{\Rhat}{\hat{R}}
\newcommand{\sigb}{\sigma_\beta}

\newcommand{\vect}[1]{\mathbf{#1}}

\begin{document}

\newpage

%%%%%%%%%%%%%%%%%%%%%%%%%%%%%%%%%%%%%%%%%%%%%%%%%%%%%%%%%%%
% Title Page
%%%%%%%%%%%%%%%%%%%%%%%%%%%%%%%%%%%%%%%%%%%%%%%%%%%%%%%%%%%

\noindent Title:

\noindent Neural integrators for decision making:  A favorable tradeoff between robustness and sensitivity
\vspace{.5cm}

\noindent Abbreviated Title:

\noindent Neural integrators for decision making
\vspace{.5cm}

\noindent Authors:

\noindent Nicholas Cain, Department of Applied Mathematics, University of Washington

\noindent Andrea K. Barreiro, Department of Applied Mathematics, University of Washington

\noindent Michael Shadlen, Department of Physiology and Biophysics, University of Washington

\noindent Eric Shea-Brown, Department of Applied Mathematics, University of Washington
\vspace{.5cm}

\noindent Corresponding Author:

\noindent Eric Shea-Brown

\noindent Department of Applied Mathematics

\noindent Box 325420

\noindent Seattle, WA 98195-2420

\noindent etsb@uw.edu

\vspace{.5cm}

\noindent Number of pages: 30

\noindent Number of figures: 16

\noindent Number of tables: 1

\noindent Number of words (Abstract): 226

\noindent Number of words (Introduction):  476

\noindent Number of words (Discussion): 1467
\vspace{.5cm}

%\noindent Contents of online supplement:
%
%\noindent In the online supplement, we include additional studies of compatibility of the robust integrator models with behavioral data from two-alternative decision making tasks.  Specifically, we compare RT histograms separated into correct and incorrect trials, and report quantitative fits with controlled duration accuracy measurements reported in \citetext{Kiani:2008ee}.  We also report ramping, trial-averaged firing rate functions that qualitatively agree with those reported in \citetext{Roitman:2002wr}, and decision-triggered stimulus traces in agreement with the data of \citetext{Kiani:2008ee}.  We also review how the simplified integrator model used in this paper can be related to the dynamics of coupled neural populations.
%\vspace{.5cm}

\noindent Conflict of Interest: None

\paragraph{Acknowledgements:}
This research was supported by a Career Award at the Scientific Interface from the Burroughs-Wellcome Fund (ESB), the Howard Hughes Medical Institute, the National Eye Institute Grant EY11378, and the National Center for Research Resources Grant RR00166 (MS), by a seed grant from the Northwest Center for Neural Engineering (ESB and MS), and by NSF Teragrid allocation TG-IBN090004.

\newpage

\begin{abstract}
	A key step in many perceptual decision tasks is the integration of sensory inputs over time, but fundamental questions remain about how this is accomplished in neural circuits.  One possibility is to balance decay modes of membranes and synapses with recurrent excitation.  To allow integration over long timescales, however, this balance must be precise; this is known as the fine tuning problem.  The need for fine tuning can be overcome via a ratchet-like mechanism, in which momentary inputs must be above a preset limit to be registered by the circuit.  The degree of this ratcheting embodies a tradeoff between sensitivity to the input stream and robustness against parameter mistuning.
	
	The goal of our study is to analyze the consequences of this tradeoff for decision making performance.  For concreteness, we focus on the well-studied random dot motion discrimination task.  For stimulus parameters constrained by experimental data, we find that loss of sensitivity to inputs has surprisingly little cost for decision performance.  This leads robust integrators to performance gains when feedback becomes mistuned.  Moreover, we find that substantially robust and mistuned integrator models remain consistent with chronometric and accuracy functions found in experiments.   We explain our findings via sequential analysis of the momentary and integrated signals, and discuss their implication:  robust integrators may be surprisingly well-suited to subserve the basic function of evidence integration in many cognitive tasks.

%via sequential analysis of the signals accumulated by different circuit models.

%analyzing the momentary statistics of inputs to the circuits, and how these statistics are correlated across time.

\end{abstract}

%%%%%%%%%%%%%%%%%%%%%%%%%%%%%%%%%%%%%%%%%%%%%%%%%%%%%%%%%%%%
% Introduction
%%%%%%%%%%%%%%%%%%%%%%%%%%%%%%%%%%%%%%%%%%%%%%%%%%%%%%%%%%%%
\section{Introduction}

Many decisions are based on the balance of evidence that arrives at different points in time.  This process is quantified via simple perceptual discrimination tasks, in which the momentary value of a sensory signal carries negligible evidence but correct responses arise from summation of this signal over the duration of a trial.  At the core of such decision making must lie neural mechanisms that integrate signals over time~\cite{Gold:2007fo,Wang:2008jx,Bogacz:2006fj}.  The function of these circuits is intriguing, because perceptual decisions develop over hundreds of milliseconds to seconds, while individual neuronal and synaptic activity often decays on timescales of several to tens of milliseconds --  a difference of at least an order of magnitude.  A mechanism that bridges this gap is feedback connectivity tuned to balance -- and hence cancel -- inherent voltage leak and synaptic decay~\cite{springerlink:10.1007/BF00320393,Usher:2001vq}.

The tuning of recurrent connections to achieve this balance presents a challenge~\cite{Seung:1996va,Seung:2000uv}, illustrated in Figure~\ref{fig:Energy_Surface}(A) via motion of a ball on an energy surface.  Here, the ball position $E(t)$ represents the total activity of a circuit (relative to a baseline marked 0); momentary sensory input perturbs $E(t)$ to increase or decrease.  If decay dominates (upper-right), then $E(t)$ always has a tendency to ``roll back'' to baseline values, thus forgetting accumulated sensory input.  Conversely, if feedback connections are in excess, then activity will grow away from the baseline value (center).  If balance is perfectly achieved via fine-tuning, (left) temporal integration can occur.  That is,  inputs can then smoothly perturb network activity back and forth, so that the network state at any given time represents the time-integral of past inputs.

\citetext{Koulakov:2002kx} proposed an alternate model:  a ratchet-like accumulator, equivalent to movement along a scalloped energy surface (Figure~\ref{fig:Energy_Surface}(A), bottom)~\cite{Pouget:2002va}.  Importantly, even without finely-tuned connectivity, network states can hold prior values without decay or growth, allowing integration of inputs over time.  Thus, this mechanism is called a {\it robust integrator}.  Energy wells can be spaced arbitrarily close together while maintaining their depth, so that the robust integrator can represent a practically continuous range of values.  However, the energy wells imply a minimum input strength to transition between adjacent states, with inputs below this limit effectively ignored.

The two models just introduced present a tradeoff between robustness to parameter mistuning and sensitivity to inputs.  Here, we ask how this tradeoff impacts behavioral performance in perceptual decision making.  Focusing on the moving dots task~\cite{Shadlen:1996wj,Roitman:2002wr}, enables us to constrain model parameters to known physiology and behavior. Our aim is to establish whether or not the robust integrator model is consistent with known data, and to assess the performance benefits, if any, that it affords when network parameters cannot be fine-tuned.
	%FIG*FIG*FIG*FIG*FIG*FIG*FIG*FIG*FIG*FIG*FIG*FIG*FIG*FIG*FIG*FIG*FIG*FIG*FIG*FIG
\begin{figure}
	\makebox[6.5in][l]{(A)\hspace{3in}(B)}
	\begin{minipage}[b]{3in}
		\centering
		\includegraphics[width=3in]{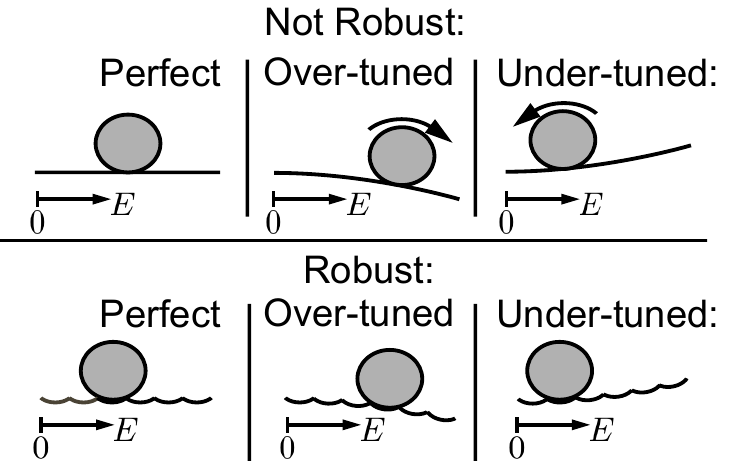}
	\end{minipage}
	\begin{minipage}[b]{3in}
		\includegraphics[width=3in]{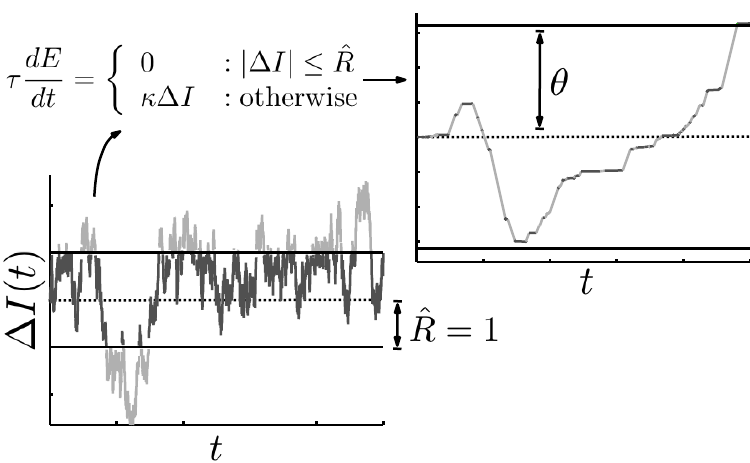}
	\end{minipage}
	\caption{Schematic of neural integrator models. (A) Visualizing integration via an energy surface \protect\cite{Pouget:2002va}.  A robust integrator can ``fixate" at a range of discrete values, indicated by a sequence of potential wells, despite mistuning of circuit feedback.  Without these wells (the non-robust case), activity in a mistuned integrator would either exponentially grow or decay, as in the top panels.  Perturbing the robust integrator from one well to the next, however, requires sufficiently strong momentary input. (B) As a consequence, low-amplitude segments in the input signal $\Delta I(t)$, below a robustness limit $R$, are not accumulated by a robust integrator:  only the high-amplitude segments are.  The piecewise-defined differential Equation~\eqref{eq:chopModel} captures this robustness behavior, resulting in the accumulated activity shown, and may be related to, e.g., a detailed bistable-subpopulation model.  A decision is expressed
	% --- here, in favor of the preferred alternative ---
when the accumulated value $E(t)$ crosses the decision threshold $\theta$.}
	\label{fig:Energy_Surface}
\end{figure}
%FIG*FIG*FIG*FIG*FIG*FIG*FIG*FIG*FIG*FIG*FIG*FIG*FIG*FIG*FIG*FIG*FIG*FIG*FIG*FIG

%%%%%%%%%%%%%%%%%%%%%%%%%%%%%%%%%%%%%%%%%%%%%%%%%%%%%%%%%%%%
% Materials and Methods
%%%%%%%%%%%%%%%%%%%%%%%%%%%%%%%%%%%%%%%%%%%%%%%%%%%%%%%%%%%%
\section{Materials and methods}

%%%%%%%%%%%%%%%%%%%%%%%%%%%%%%%%%%%%%%%%%%%%%%%%%%%%%%%%%%%%
%  Model Design
%%%%%%%%%%%%%%%%%%%%%%%%%%%%%%%%%%%%%%%%%%%%%%%%%%%%%%%%%%%%
\subsection{Model and task overview}

To explore the consequences of the robust integrator mechanism for decision performance, we begin by constructing a two-alternative decision making model similar to that proposed by \citetext{Mazurek:2003cm}.  For concreteness, we concentrate on the forced choice motion discrimination task~\cite{Roitman:2002wr,Mazurek:2003cm,Gold:2007fo,Churchland:2008wg,Shadlen:1996wj,Shadlen:2001ve}.  Here, subjects are presented with a field of random dots, of which a subset move coherently in one direction; the remainder are relocated randomly in each frame.  The task is to correctly choose the direction of coherent motion from two alternatives (i.e., left vs. right).

\begin{figure}
	\centering
	\includegraphics[width=6in]{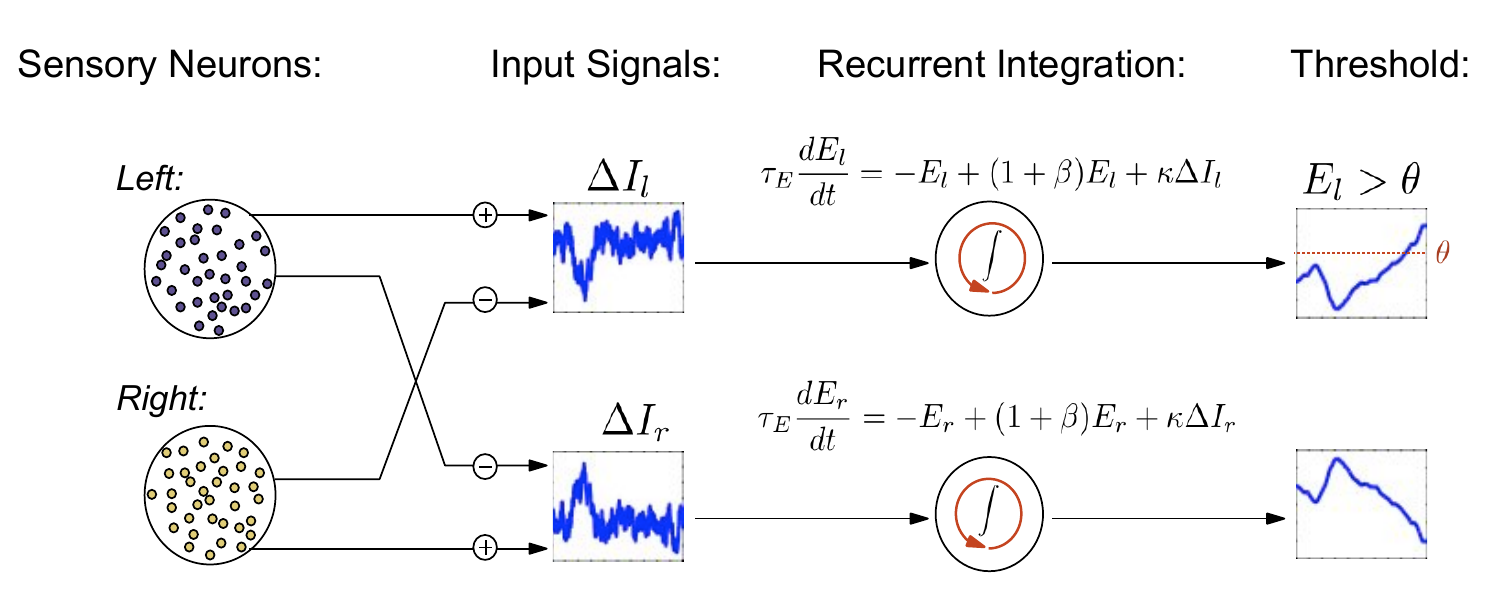}
	\caption{Overview of  model setup.  Simulations of sensory neurons and neural recordings are used to define the left and right inputs $\Delta I_l(t)$, $\Delta I_r(t)$ to neural integrators (see text).  These inputs are modeled by gaussian (OU) processes, which capture noise in the encoding of the motion strength by each pool of spiking neurons.  See Equations \eqref{eq:MT_statistics1}-\eqref{eq:quantities} for definition of input signals.  Similar to \protect\citetext{Mazurek:2003cm}, the activity levels of the left and right integrators $E_l(t)$ and $E_r(t)$ encode accumulated evidence for each alternative.  In the reaction time task, $E_l(t)$ and $E_r(t)$ race to thresholds in order to determine choice on each trial.   In the controlled duration task the choice is made in favor of the integrator with higher activity at the end of the stimulus presentation.}
\label{fig:modelOverview}
\end{figure}

As in \citetext{Mazurek:2003cm} (see also \citetext{Smith:2010wq}), we first simulate a population of neurons that represent the sensory input to be integrated over time.  This  population is a rough model of cells in extrastriate cortex (Area MT) which encode momentary information about motion direction~\cite{Britten:1993wv,Britten:1992wx,Salzman:1992wg}.  We pool spikes from model MT cells that are selective for each of the two possible directions into separate streams, labeled according to their preferred ``left" and ``right" motion selectivity:  see Figure~\ref{fig:modelOverview}.

Two corresponding integrators then accumulate the difference between these streams,  left-less-right or vice-versa.  Each integrator therefore accumulates the evidence for one alternative over the other.
Depending on the task paradigm, different criteria may be used to terminate accumulation and give a decision.
In the \textit{reaction time} task, accumulation continues until activity crosses a decision threshold:  if the leftward evidence integrator reaches threshold first, a decision that overall motion favored the leftward alternative is registered.

Accuracy is defined as the fraction of trials that reach a correct decision.  Speed is measured by the time taken to cross threshold starting from stimulus onset.  Reaction Time (RT) is then defined as the time until threshold (decision time) plus 350 ms of non-decision time, accounting for other delays that add to the time taken to select an alternative (e.g. visual latencies, or motor preparation time, cf.~\cite{Mazurek:2003cm,Luce:1986vp}).  The exact value of this parameter was not critical to our results.
Task difficulty is determined by the fraction of coherently moving dots $C$~\cite{Britten:1992wx,Mazurek:2003cm,Roitman:2002wr}. Accuracy and RT across multiple levels of task difficulty define the accuracy and chronometric functions in the reaction time task, and together can be used to assess model performance. When necessary, these two numbers can be collapsed into a single metric, such as the reward per unit time or \textit{reward rate}.

In a second task paradigm, the \textit{controlled duration} task, motion viewing duration is set in advance by the experimenter.  A choice is made in favor of the integrator with greater activity at the end of the stimulus duration.  Here, the only measure of task performance is the accuracy function.

%FIG*FIG*FIG*FIG*FIG*FIG*FIG*FIG*FIG*FIG*FIG*FIG*FIG*FIG*FIG*FIG*FIG*FIG*FIG*FIG*F
%FIG*FIG*FIG*FIG*FIG*FIG*FIG*FIG*FIG*FIG*FIG*FIG*FIG*FIG*FIG*FIG*FIG*FIG*FIG*FIG*F

%%%%%%%%%%%%%%%%%%%%%%%%%%%%%%%%%%%%%%%%%%%%%%%%%%%%%%%%%%%%
%  Model Input
%%%%%%%%%%%%%%%%%%%%%%%%%%%%%%%%%%%%%%%%%%%%%%%%%%%%%%%%%%%%
\subsection{Sensory input}

%FIG*FIG*FIG*FIG*FIG*FIG*FIG*FIG*FIG*FIG*FIG*FIG*FIG*FIG*FIG*FIG*FIG*FIG*FIG*FIG*
\begin{figure}
	\centering
	\includegraphics[width=6in]{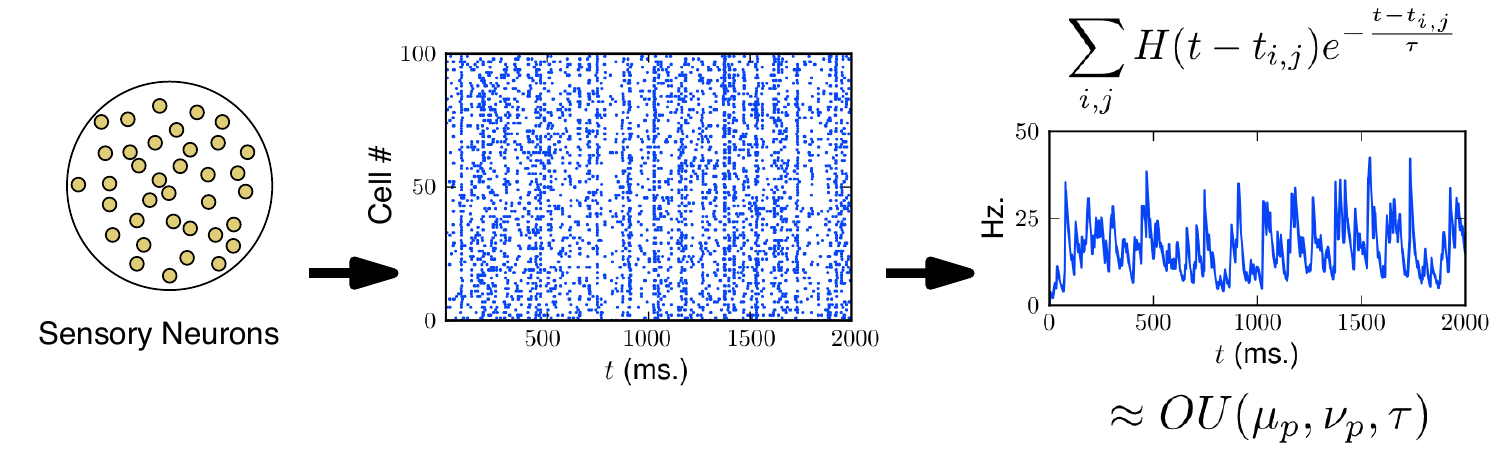}
	\caption{Construction of gaussian (OU) processes to represent fluctuating, trial-by-trial firing rate of a pool of weakly correlated MT neurons \protect\cite{Bair:2001wm,Zohary:1994uv}.  As in \protect\citetext{Mazurek:2002eo}, these motion sensitive neurons provide direct input to our model integrator circuits.  Simulated spike trains from weakly-correlated, direction selective pools of neurons are shown as a rastergram.  All spikes prior to time $t$ -- a sum over the $j^{th}$ spike from the $i^{th}$ neuron, for all $i$ and $j$ -- are convolved with an exponential filter, and then summed to create a continuous stochastic output (right); here, $H(t)$ is the Heaviside function.  We approximated this output by a simpler gaussian (OU) process in order to simplify  numerical and analytical computations that follow.   %The steady-state mean, variance, and time constant of this process provide the sufficient statistics for an Orstein-Uhlenbeck (OU) process; we approximated complicated spiking model }
}
	\label{fig:rasterBin}
\end{figure}
%FIG*FIG*FIG*FIG*FIG*FIG*FIG*FIG*FIG*FIG*FIG*FIG*FIG*FIG*FIG*FIG*FIG*FIG*FIG*FIG*

We now describe in detail the signals that are accumulated by the integrators corresponding to the ``left" and ``right" alternatives.  First, we model the pools of leftward or rightward direction-selective sensory (MT) neurons as $N=100$ weakly correlated spiking cells (Pearson's correlation $\rho=.11$~\cite{Zohary:1994uv,Bair:2001wm}); see Figure~\ref{fig:rasterBin}.  Specifically,  as in \citetext{Mazurek:2002eo}, each neuron is modeled via an unbiased random walk to a spiking threshold; the random walks of neurons in the same pool are correlated.  Increasing the variance of each step in the random walk increases the firing rate of each model neuron; it was therefore chosen at each coherence value to reproduce the linear relationship between coherence $C$ and mean firing rate $\mu_{l,r}$ of the left and right selective neurons observed in MT recordings:
\begin{equation}
	\mu_{l,r}(C) = r_0 + b_{l,r}C \; \; .
	\label{eq:MT_statistics1}
\end{equation}
Here the parameters $r_0$, $b_l$, and $b_r$ are derived from firing rates observed across a range of coherences~\cite{Britten:1993wv}.  If evidence favors the left alternative, $b_l=.4$ and $b_r=-.2$; if the right alternative is favored, these values are exchanged.

Next, the output of each spiking pool was aggregated.  Each spike emitted from a neuron in the  pool was convolved with an exponential filter with time constant $20 $ ms. This is intended as an approximate model of the smoothing effect of synaptic transmission. These smoothed responses were then summed to form a single stochastic process for each pool (see Figure~\ref{fig:rasterBin}, right). %  A linear relationship between the mean and variance in the resulting process was observed across coherence values, with a proportionality constant of $4.5$.
% THIS SENT COMMENTED OUT IN V19 ... could confuse, and not clearly necessary for what follows (since fit)

%We note that it is also possible to derive an analogous relation for a simplified Poisson spiking model; details are provided in the online supplement.
%\begin{equation}
%	\nu_{l,r}(C) = k_{\nu}\mu_{l,r}(C) \label{eq:MT_statistics3}
%\end{equation}

We then approximated the smoothed output of each spiking pool by a simpler stochastic process that captures the mean, variance, and temporal correlation of this output as a function of dot coherence.   We used gaussian processes $I_l(t)$ and $I_r(t)$ for the rightward- and leftward-selective pools (See Figure~\ref{fig:rasterBin}).  Specifically, we chose Ornstein-Uhlenbeck (OU) processes, which are continuous gaussian process generated by the stochastic differential equations
\beq
	dI_{l,r} = \frac{\mu_{l,r}(C)-I_{l,r}}{\tau}dt + \sqrt{\frac{2\nu_{l,r}(C)}{\tau}}dW_t \; \;
\eeq
with mean $\mu_{l,r}(C)$ as dictated by Equation \ref{eq:MT_statistics1}.  The variance $\nu_{l,r}(C)$ and timescale $\tau$ were chosen to match the steady-state variance and autocorrelation function of the smoothed spiking process.  As we will see, this timescale plays an important role in determining the decision making performance of robust integrators.

Our construction so far accounts for variability in output from left vs. right direction selective neurons.  We now incorporate an additional noise source into the output of each pool.  These noise terms ($\eta_l(t)$ and $\eta_r(t)$, respectively) could approximate, for example, neurons added to each pool that are nonselective to direction.
% represents elements of each pool that are less selective to the stimulus direction (i.e., ,  of stimulus motion. These could be spikes from neurons selective for directions other than the leftward or rightward alternatives, or  inputs into the integrator from regions that are unrelated to the perceptual task. We represent this noise source as an additional stochastic term added onto the output of each sensory pool.
Each noise source is modeled as an independent OU process with mean 0, timescale 20 ms as above, and a strength (variance) $\nu_{\gamma}/2$.  This noise strength is a free parameter that we vary to match behavioral data (see "A robust integrator circuit" and Figure~\ref{fig:naivePsychoCrono}).  We note that previous studies also found that performance based on the direction-sensitive cells alone can be more accurate than behavior, and therefore incorporated variability in addition to the output of ``left" and ``right" direction selective MT cells~\cite{Shadlen:1996wt,Mazurek:2003cm,Cohen:2009tt}.

%(e.g., pooling noise was used in \citetext shad, doubly stochastic processes in \citetext{maz}, and pooling correlation in \citetext{cohen}).

Finally, the signals that are accumulated by the left and right neural integrators are constructed by differencing the outputs of the two neural pools:
\begin{eqnarray}
	\Delta I_{l}(t) & = &  \left[I_{l}(t) + \eta_l(t)\right]  -  \left[I_{r}(t) +\eta_r(t)\right]  \nonumber \\
	 \Delta I_{r}(t) & = &-\Delta I_{l}(t) \label{eq:quantities} \; \;.
\end{eqnarray}

%where the additional unbiased additive noise $I_{\gamma}(t)$ is given by
%\beqn
%	dI_{\gamma}  =  -\frac{I_{\gamma}}{\tau}dt + \sqrt{\frac{2\nu_{\gamma}}{\tau}}dW_t \;.
%\eeqn
%
%Previous studies  %The amount of additive noise $\nu_\gamma$ necessary to match model performance with known psychophysics (Fig. \ref{fig:naivePsychoCrono}) in our model was similar to that in \citetext{Shadlen:1996wt}.

\subsection{Neural integrator circuit and feedback mistuning}

%The input signals that are accumulated by the leftward and rightward integrators have already addressed two sources of variability, intrinsic stochastic fluctuation and additive noise.

A central focus of our paper is variability in the relative tuning of recurrent feedback vs. decay in an integrator circuit.  Below, we will introduce the {\it mistuning parameter} $\beta$, which determines the extent to which feedback and decay fail to perfectly balance.
We first define the dynamics of the integrator circuit on which our studies are based.  This is described by the firing rates $E_{l,r}(t)$ of integrators that receive outputs from left-selective or right-selective pools $\Delta I_{l,r}(t)$ respectively.  The firing rates $E_{l,r}(t)$ increase as evidence for the corresponding task alternative is accumulated over time:
	%EQN*EQN*EQN*EQN
	\begin{equation}
		\tau_E \frac{dE_{l,r}}{dt} = -E_{l,r} + (1+\beta)E_{l,r} + \kappa \Delta I_{l,r}(t).
		\label{eq:naive}
	\end{equation}
	%EQN*EQN*EQN*EQN
	
\noindent The three terms in this equation account for leak, feedback excitation, and the sensory input (scaled by a weight $\kappa$), respectively.  When the mistuning parameter $\beta=0$, leak and self-excitation exactly cancel; we describe such an integrator as \textit{perfectly tuned}, while an integrator with $\beta \neq 0$ is said to be \textit{mistuned}.  Imprecise feedback tuning is modeled by randomly setting $\beta$ to different values from trial to trial (but constant during a given trial), with a mean value $\bar{\beta}$ and a precision given by a standard deviation $\sigma_{\beta}$.  We assume that $\bar{\beta}=0$ for most of the study.  Thus the spread of $\beta$, which we take to be gaussian, represents the intrinsic variability in the balance between circuit-level feedback and decay.  Perfect tuning corresponds to $\sigma_{\beta}=\bar{\beta}=0$, while $\sigma_{\beta}\neq0$ or $\bar{\beta}\neq0$ corresponds to a mistuned integrator.
 Finally, we set initial activity in the integrators to zero ($E_{l,r}(0)=0)$, and impose reflecting boundaries at $E_r=0$, $E_l = 0$ (as in, e.g., \citetext{Smith:2004wu}) so that firing rates never become negative.

\subsection{A robust integrator circuit}
\label{sec:robustNeuralIntegrator}

A robust integrator can be constructed from a series of bistable subpopulations, which sequentially activate in order to represent the accumulated evidence~\cite{Koulakov:2002kx,Nikitchenko:2008wt}.  The many equations that describe the evolution of these systems can be closely approximated with reduced models, as demonstrated in ~\citetext{Goldman:2003ge}.  We derived a single piecewise-defined differential equation model that approximates the dynamics of a robust integrator constructed from bistable pools.  

All subsequent results are based on this simplified model, which captures the essence of the robust integration computation:
\begin{equation}
	\tau_E \frac{dE_{l,r}}{dt} = \left\{
	\begin{array}{ll}
		0  & :  \left|\beta E_{l,r} + \kappa \Delta I_{l,r} \right| \leq R \\
		\beta E_{l,r} + \kappa \Delta I_{l,r}  & :  \text{otherwise}
	\end{array}
	\right.
	\label{eq:chopModel}
\end{equation}
The first line represents the series of potential wells discussed in the Introduction (see Figure~\ref{fig:Energy_Surface}):  if the sum of the mistuned integrator feedback and the input falls below the robustness limit $R$, the activity of the integrator remains fixed.  If this summed input exceeds $R$, the activity evolves as for the non-robust integrator in Equation~\eqref{eq:naive}.  To interpret the robustness limit $R$, it is convenient to normalize by the standard deviation of the input signal:   %Specifically, consider the preferred input signal $\Delta I_p$; the units of $R$ are clearly the same as the mean $\mu_{p}$ - $\mu_{n}$, and standard deviation of an input signal. Because $\Delta I_p$ is a stationary process, it convenient to define this fraction as:
\begin{equation}
	\hat{R} = \frac{R}{STD\left[\Delta I_{l,r}(t) \right]}.
	\label{eq:RHat}
\end{equation}
In this way, $\hat{R}$ can be interpreted in units of standard deviations of input OU process that are ``ignored" by the integrator. We note that Equation~\eqref{eq:chopModel} is similar to the effective equation derived for a different implementation of a robust integrator~\cite{Goldman:2003ge}.

To summarize, Equation~\eqref{eq:chopModel} defines a parameterized family of neural integrators, distinguished by the robustness limit $\hat{R}$.  As $\hat{R} \rightarrow 0$, the model reduces to Equation \eqref{eq:naive}.  When additionally $\beta=0$, the (perfectly tuned) integrator computes an exact integral of its input:  Equation~\eqref{eq:chopModel} then yields $E_{l,r}(t) \propto\int_0^t \Delta I_{l,r} (t') dt'$.

	\subsection{Computational methods}
	Monte Carlo simulations of Equations \eqref{eq:MT_statistics1}-\eqref{eq:chopModel} were performed with Euler-Maruyama method~\cite{Higham:2001vk}, with $dt=0.1$ ms. 	
	%The number of trials used to generate the FC and RT curves in Figures \ref{fig:Effect_Of_Chop}-\ref{fig:BetaHat} was greater than $10,000$ (FC and RT for a given threshold $\theta$ by averaging across trials).
	For a fixed choice of input statistics and threshold $\theta$, a minimum of $10,000$ trials were simulated to estimate accuracy and RT values.
	During simulations of the reaction time task, in order to prevent excessively long trials (particularly at low coherence values) a maximum simulation time was set at 10,000 ms.  At this time, if neither integrator had reached threshold, the indeterminate result was broken by a numerical ``coin flip", (this rarely occurred, as indicated by the RT histograms in Figures~\ref{fig:RTDistCons},~\ref{fig:RTDistCollapse}).
	
	%Marginalization across $\beta$ in Figures \ref{fig:BasicToMistuned}-\ref{fig:MistunedToRecovery} and \ref{fig:Effect_Of_Chop}-\ref{fig:RTDistCollapse} was accomplished by simulating across a discrete range of values.
	In simulations where $\sigma_{\beta} > 0$, results were generated across a range of $\beta$ values and then marginalized by weighting according to a normal distribution.  The range of values was chosen with no less than 19 linearly spaced points, across a range of $\pm$ 3 standard deviations around the mean $\bar{\beta}$.
		
	Reward rate values presented in ``Reward rate and the robustness-sensitivity tradeoff" are presented as maximized by varying the free parameter $\theta$; values were computed by simulating across a range of $\theta$ values. The range and spacing of these values were chosen dependent on the values of $\hat{R}$ and ${\beta}$ for the simulation; the range was adjusted to capture the relative maximum of reward rate as a function of $\theta$, while the spacing was adjusted to find the optimal $\theta$ value with a resolution of $\pm$ $0.1.$
	
%In order to match our models to behavioral data, must fix the free parameters $\theta$ and $\mu_{\gamma}$, and the robustness parameter $R$.  so as to minimize the sum-squared error in data (from \cite{Roitman:2002wr}) vs. model psychometric and chronometric curves. Model curves were evaluated for values of $\theta$ and $\hat{R}$ on a discrete grid with a resolution of 0.1. When data were needed that were between simulated values, linear interpolation was used to approximate the corresponding FC and RT values.

Values of $\theta$ and $\nu_\gamma$ in the table included in Figure~\ref{fig:naivePsychoCrono} were chosen to best match accuracy and chronometric functions to behavioral data reported in \citetext{Roitman:2002wr}.  This was accomplished by minimizing the sum-squared error in data vs. model accuracy and chronometric curves across a discrete grid of $\theta$ and $\nu_\gamma$ values, with a resolution of 0.1.  When data between simulated values were needed, linear interpolation was used to approximate the corresponding accuracy and RT values.

	Autocovariance functions of integrator input, presented later, %plotted in Figure \ref{fig:FDPrediction} (B)
	were computed by simulating an Ornstein-Uhlenbeck process using the exact numerical technique in \citetext{Gillespie:1996ve} with $dt=0.1$ ms, to obtain a total of $2^{27}$ sample values. Sample values of the process less than the specified robustness limit $\hat{R}$ were set to 0, and the autocovariance function was computed using standard Fourier transform techniques.

	Simulations were performed on NSF Teragrid clusters.

%%%%%%%%%%%%%%%%%%%%%%%%%%%%%%%%%%%%%%
%%%%%%%%%%%%%%%%%%%%%%%%%%%%%%%%%%%%%%
%%%%%%%%%%%%%%%%%%%%%%%%%%%%%%%%%%%%%%
%%%%%%%%%%%%%%%%%%%%%%%%%%%%%%%%%%%%%%
%%%%%%%%%%%%%%%%%%%%%%%%%%%%%%%%%%%%%%
%%%%%%%%%%%%%%%%%%%%%%%%%%%%%%%%%%%%%%
%%%%%%%%%%%%%%%%%%%%%%%%%%%%%%%%%%%%%%

\begin{figure}
	\centering
	\includegraphics[width=3in]{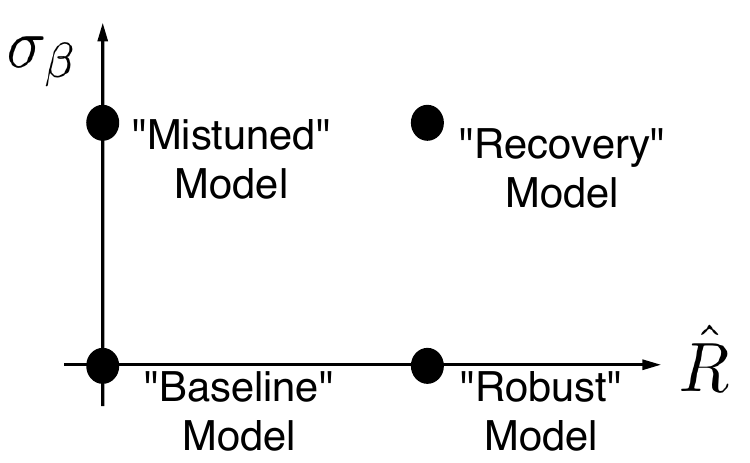}
	\caption{Parameter space view of four integrator models, with different values of the robustness limit $\hat{R}$ and feedback mistuning $\sigma_{\beta}$.  The impact of transitioning from one model to another by changing parameters is either to enhance or diminish performance, or to have a neutral effect (see text).}
	\label{fig:2ParamOverview}
\end{figure}

\section{Results}

\subsection{How do robustness and mistuning affect decision speed and accuracy?}
\label{dot sec:MistuningAndRobustness}

In the Methods, we define a general neural integrator model (Equation~\eqref{eq:chopModel}) that accumulates signals representing the output of motion sensitive neurons (Equation~\ref{eq:quantities}).  The integrator model includes two key parameters.  The first is $\beta$, which represents the  mistuning of feedback from a value that perfectly balances decay; the extent of this mistuning is measured by $\sigma_\beta$, the standard deviation of $\beta$ from the ideal value $\beta=0$.   The second is the robustness limit $\Rhat$.  We emphasize twin effects of $\hat R$:  as $\hat{R}$ increases, the integrator becomes able to produce a range of graded persistent activity for ever-increasing levels of mistuning (see Figure~\ref{fig:Energy_Surface} (A), where $\hat R$ corresponds to the depth of energy wells).  This prevents runaway increase or decay of activity when integrators are mistuned; intuitively, this might lead to better performance on sensory accumulation tasks.  At the same time, as $\hat{R}$ increases integrator activity remains fixed even for increasingly strong positive or negative momentary input $\Delta I_{l,r} $ (see Figure~\ref{fig:Energy_Surface} (B), where $\hat R$ specifies a limit within which inputs are ``ignored"). Such sensitivity loss should lead to worse performance.   %In effect, this mechanism inactivates the integration circuit for inputs below the robustness limit (in magnitude).
This implies a fundamental tradeoff between competing effects: (1) one would prefer to not ignore relevant input stimulus, favoring small $\hat{R}$, and (2) one would prefer an integrator robust to mistuning, favoring large $\hat{R}$.

Figure~\ref{fig:2ParamOverview} gives a schematic of how the two model parameters, $\sigma_{\beta}$ and $\hat{R}$,  define a plane of possible integrator models.  Here, we explore decision performance in four different cases arranged in this plane.  By contrasting integrators with different values of the robustness limit $\Rhat$, we can assess how the fundamental tradeoff plays out, to either improve or degrade decision making performance.

In order to assess this performance, we consider relationships between decision speed and accuracy in both controlled duration and reaction time tasks.  In the controlled duration task, we simply vary the stimulus presentation duration, and plot accuracy vs. experimenter-controlled stimulus duration.
In the reaction time task, we vary the decision threshold $\theta$ --- treated as a free parameter --- over a range of values, thus tracing out the speed accuracy curve for all possible pairs of speed and accuracy values.  Here, speed is measured by reaction time (RT), the latency between the onset of stimulus and crossing of the decision threshold.  For both cases, we use a single representative dot coherence (C=12.8 in Equation \ref{eq:MT_statistics1}); results are qualitatively similar for other coherence values (data not shown); slightly (approx. $25\%$) lower robustness limits are required at the lowest dot coherence of $C=3.2$.

%FIG*FIG*FIG*FIG*FIG*FIG*FIG*FIG*FIG*FIG*FIG*FIG*FIG*FIG*FIG*FIG*FIG*FIG*FIG*FIG*
\begin{figure}
	\centering
	\includegraphics[width=6in]{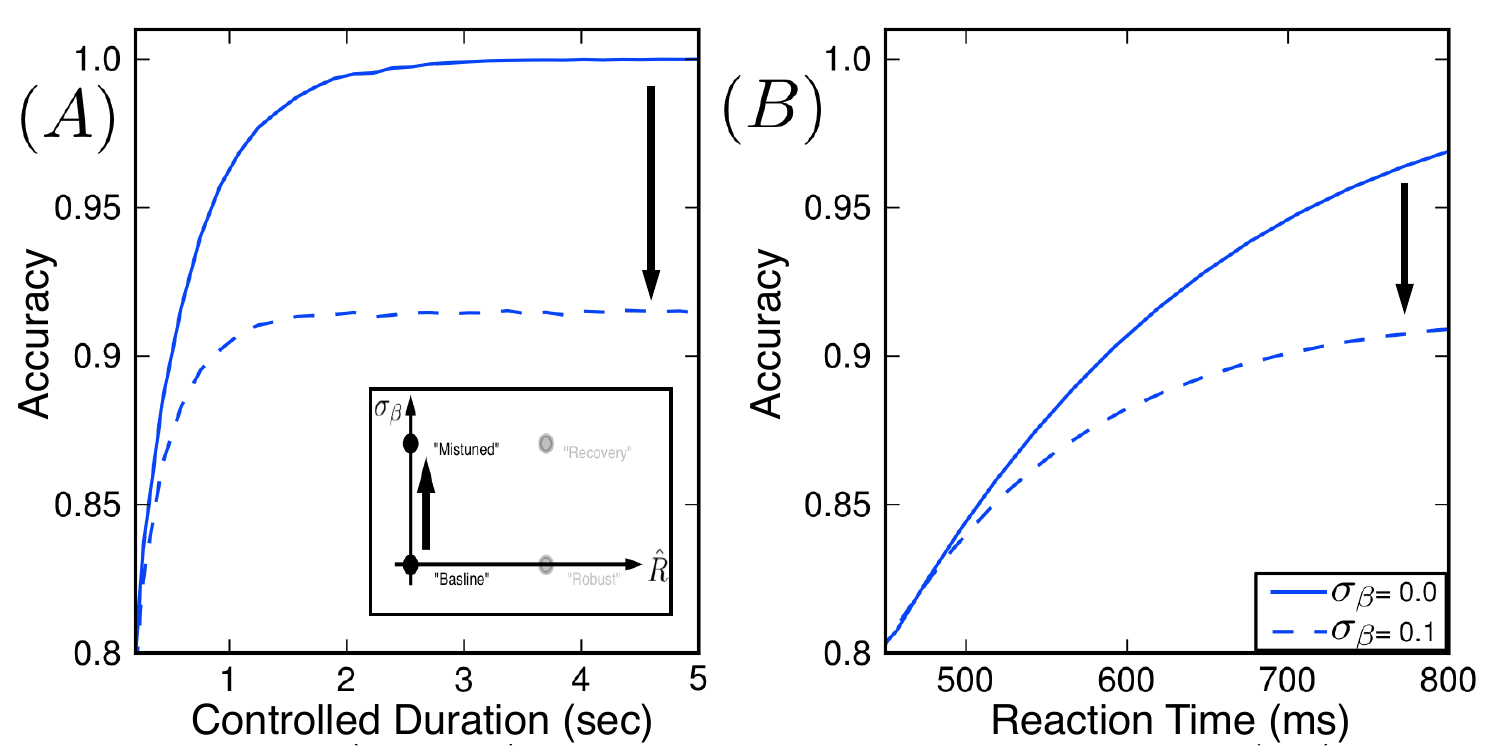}
	\caption{Mistuned feedback diminishes decision performance.  (Inset) In both figures we consider a move in parameter space from the ``baseline" model to the ``mistuned" model by changing $\sigma_{\beta}=0 \rightarrow 0.1$ (A) In the controlled duration task, accuracy is lower for the ``mistuned" model (dashed line) than for the ``baseline" model (solid line) at every trial duration $T$, indicating a loss of performance  when $\sigma_{\beta}$ increases. (B) In the reaction time task, we plot the curve of all (RT, accuracy) pairs attained by varying the decision threshold $\theta$ (see text).  Once again, accuracy is diminished by mistuning.}
	\label{fig:BasicToMistuned}
\end{figure}
%FIG*FIG*FIG*FIG*FIG*FIG*FIG*FIG*FIG*FIG*FIG*FIG*FIG*FIG*FIG*FIG*FIG*FIG*FIG*FIG*

We first study a case we call the ``baseline" model, for which there is no mistuning or robustness:  $\sigb = \Rhat=0.$   Speed accuracy plots for this model are shown as a solid line in Figs.~\ref{fig:BasicToMistuned}(A) and (B), for the controlled duration and reaction time tasks respectively.  We compare the ``baseline" model with the ``mistuned" model, for which the feedback parameter has a standard deviation of
$\sigb =0.1$ ($10\%$ of the mean feedback) and robustness $\Rhat=0$ remains unchanged.  In the controlled duration task (Figure~\ref{fig:BasicToMistuned}(A)) we observe that mistuning diminishes accuracy by as much as $10\%$, and this effect is sustained even for arbitrarily long viewing windows (cf.~\cite{Usher:2001vq,Bogacz:2006fj}).  The reaction time task (Panel B) produces a similar effect:  for a fixed RT, the corresponding accuracy is decreased.

\begin{figure}
	\centering
	\includegraphics[width=6in]{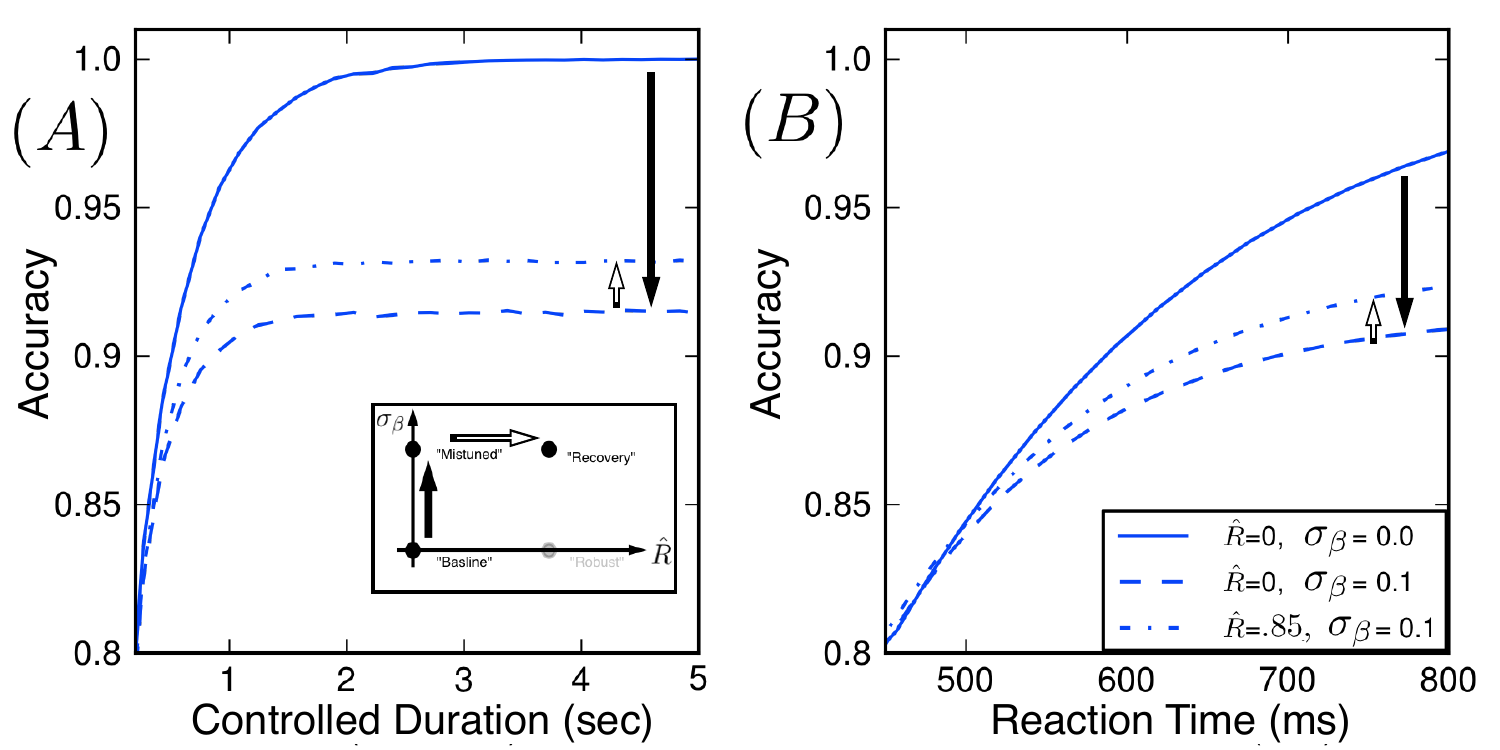}
	\caption{Increasing the robustness limit $\hat{R}$ helps recover performance lost due to feedback mistuning.  (Inset) We illustrate this by moving in parameter space from the ``mistuned" model to the ``recovery" model, by changing $\hat{R}=0 \rightarrow .85$.  The impact on decision performance is shown for both the controlled duration (A) and reaction time (B) tasks.  For each task we plot the relationship between speed and accuracy as above:  solid lines indicate the ``baseline" model, dotted the ``mistuned" model, and now dash-dotted the ``recovery" model.  We find that $\hat{R}>0$ yields a modest performance gain for the ``recovery" model in comparison with the ``mistuned" model.}
	\label{fig:MistunedToRecovery}
\end{figure}

Next we increase the robustness limit to $\hat{R}=0.85$ --- so that almost $\pm$ a standard deviation of the input stream is ``ignored" by the integrators --- while maintaining feedback mistuning.  We call this case the ``recovery" model because robustness compensates in part for the performance loss due to feedback mistuning:  the speed accuracy plots in Figure~\ref{fig:MistunedToRecovery} for the recovery case lie above those for the ``mistuned" model.  For example, at the longer controlled task durations (Panel A) and reaction times (Panel B) plotted, 20\% of the accuracy lost due to integrator mistuning is recovered via the robustness limit $\Rhat =0.85$.

%FIG*FIG*FIG*FIG*FIG*FIG*FIG*FIG*FIG*FIG*FIG*FIG*FIG*FIG*FIG*FIG*FIG*FIG*FIG*FIG*
%FIG*FIG*FIG*FIG*FIG*FIG*FIG*FIG*FIG*FIG*FIG*FIG*FIG*FIG*FIG*FIG*FIG*FIG*FIG*FIG*

%FIG*FIG*FIG*FIG*FIG*FIG*FIG*FIG*FIG*FIG*FIG*FIG*FIG*FIG*FIG*FIG*FIG*FIG*FIG*FIG*
\begin{figure}
	\centering
	\includegraphics[width=6in]{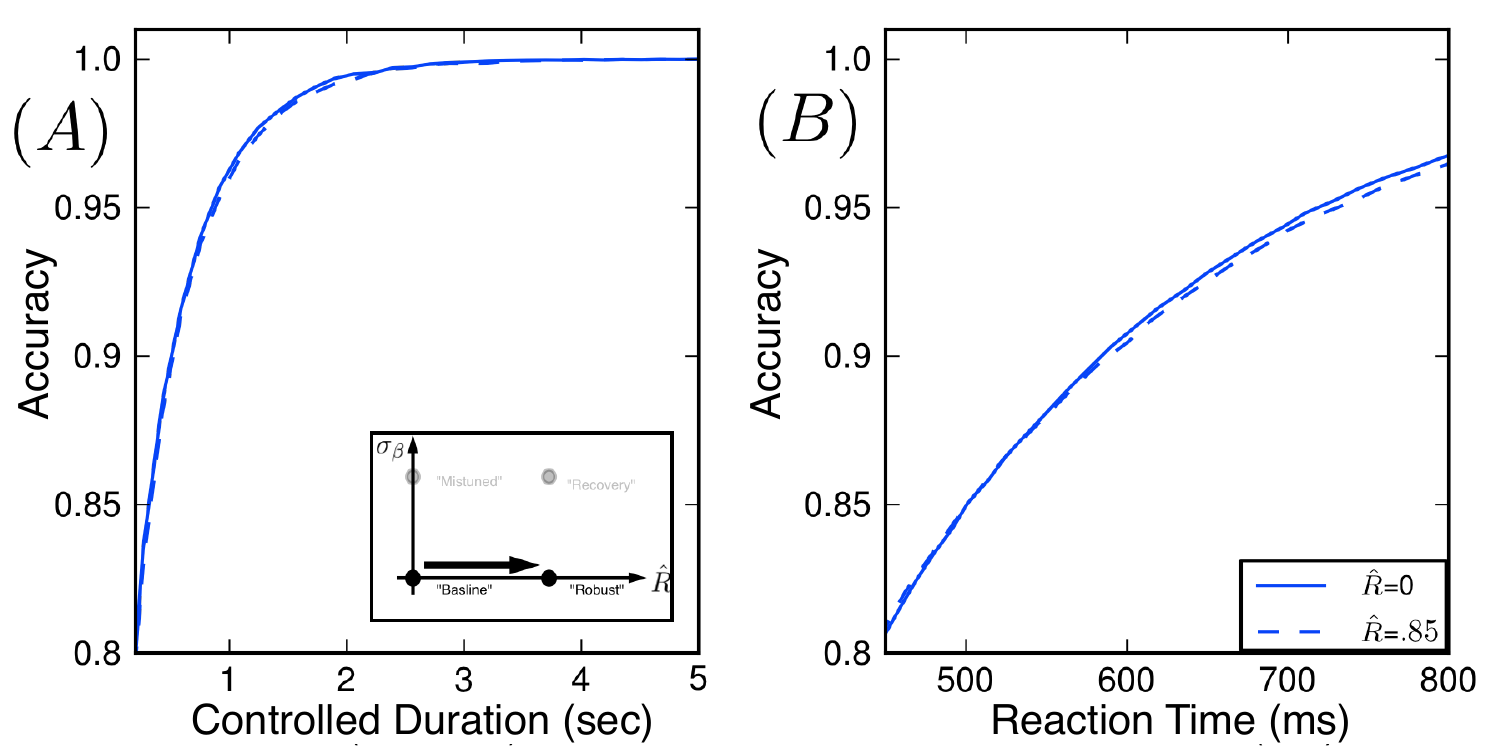}
	\caption{Increasing $\hat{R}$ alone does not compromise performance.   (Inset) We illustrate this by moving in parameter space from the ``baseline" to the ``robust" model.  For both (A) controlled duration and (B) reaction time tasks, we plot the relationship between speed and accuracy.  Solid lines give results for the ``baseline" model, and dash-dotted for the robust model.  The curves are very similar in the ``baseline" and robust cases, indicating little change in decision performance due to the robustness limit $\hat R = 0.85$. }
	\label{fig:BasicToRobust}
\end{figure}
%FIG*FIG*FIG*FIG*FIG*FIG*FIG*FIG*FIG*FIG*FIG*FIG*FIG*FIG*FIG*FIG*FIG*FIG*FIG*FIG*

Finally, we study the remaining possibility, when the robustness parameter $\Rhat$ is increased from zero in a perfectly tuned integrator ($\sigb=0$); this is the ``robust" case in Figure~\ref{fig:2ParamOverview}.  We expected performance to be substantially diminished as a consequence of lost sensitivity to inputs.  However, Figure~\ref{fig:BasicToRobust} demonstrates that this is not the case:  speed accuracy curves for $\hat{R}=0.85$ almost coincide with those for the ``baseline" case of $\hat{R}=0$. We emphasize again that because $\hat{R}$ measures ignored input in units of the standard deviation, the integrator circuit is actually not integrating the weakest $60\%$  of the input stimulus.  Given this large amount of ignored stimulus, the fact that the robust integrator produces nearly the same accuracy and speed as the ``baseline" case is surprising.  This implies that the ``robust" model can protect against feedback mistuning, without substantially sacrificing performance when feedback is perfectly tuned.

To summarize, the ratchet-like mechanism of the robust integrator appears well-suited to the decision tasks at hand.  This mechanism counteracts some of the performance lost when feedback fails to be perfectly fine-tuned.  Moreover, even when this fine-tuning is achieved, a robust integrator still performs as well as the ``baseline" case that is perfectly sensitive to the input signal.  In the next section, we begin to explain this observation by constructing several simplified models and employing results from statistical decision making theory.

% - - - - - - - - - - - - - - - - - - - - - - - - - - - - - - - - - - - - - -
\subsection{Analysis: Robust integrators and decision performance}

	\subsubsection{Controlled duration task: Discrete time analysis}	\label{sec:DiscreteAnalysisFD}

	We can begin to understand the effect of the robustness limit on decision performance by formulating a simplified version of the evidence accumulation process.  We focus first on the controlled duration task, where the analysis is somewhat simpler.
	
	Our first simplification is to consider a single accumulator $E$ which receives evidence for or against a task alternative in discrete time.  The value of $E$ on the $i^{th}$ time step, $E_i$, is allowed to be either positive or negative, corresponding to accumulated evidence favoring the leftward or rightward alternatives, respectively.  On each time step, $E_i$ increments by an independent, random value $Z_i$ with a probability density function (PDF) $f_{Z}(Z)$.  We first describe an analog of the ``baseline" model above; i.e., in the absence of robustness ($\hat R=0$).  Here, we take the increments $Z_i$ to be independent, identical, and normally distributed, with a mean $\mu>0$ (i.e. biased toward the leftward alternative; we call this the preferred alternative) and standard deviation $\sigma$:  that is, $Z_i \sim N\left( \mu, \sigma^2 \right)$.  After the $n^{th}$ step, we have
\begin{equation*}
	E_n =  \sum_{i=1}^n  Z_i \;.
\end{equation*}

In the controlled duration task, a decision is rendered after a fixed number of time steps $N$, (i.e. $n=N$) and a correct decision (i.e., in favor of the preferred alternative) occurs when $E_N>0$.	By construction, $E_n \sim N\left(n\mu,n\sigma^{2}\right)$, which implies that accuracy ($Acc$) can be computed as a function of the signal-to-noise ratio (SNR) $s=\frac{\mu}{\sigma}$ of a sample:
	\begin{equation}
		Acc = \int_{0}^{\infty}\frac{1}{\sqrt{2\pi N\sigma^{2}}}e^{-\frac{\left(x-N\mu\right)^{2}}{2N\sigma^{2}}}dx=\frac{1+\text{Erf}\left(\sqrt{\frac{N}{2}}s\right)}{2} \; .
		\label{eq:FDPsycho}
	\end{equation}

	%FIG*FIG*FIG*FIG*FIG*FIG*FIG*FIG*FIG*FIG*FIG*FIG*FIG*FIG*FIG*FIG*FIG*FIG*FIG*FIG
\begin{figure}
	\makebox[6in][l]{(A)\hspace{3in}(B)}
	\begin{minipage}[b]{3in}
		\centering
		\includegraphics[width=3in]{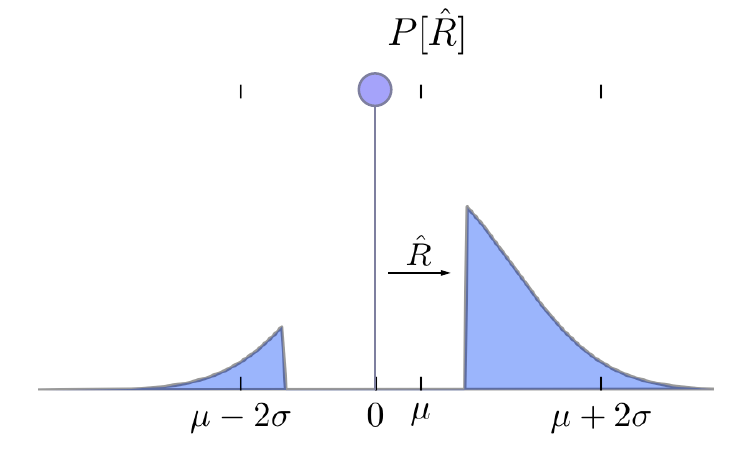}
	\end{minipage}
	\begin{minipage}[b]{3in}
		\includegraphics[width=3in]{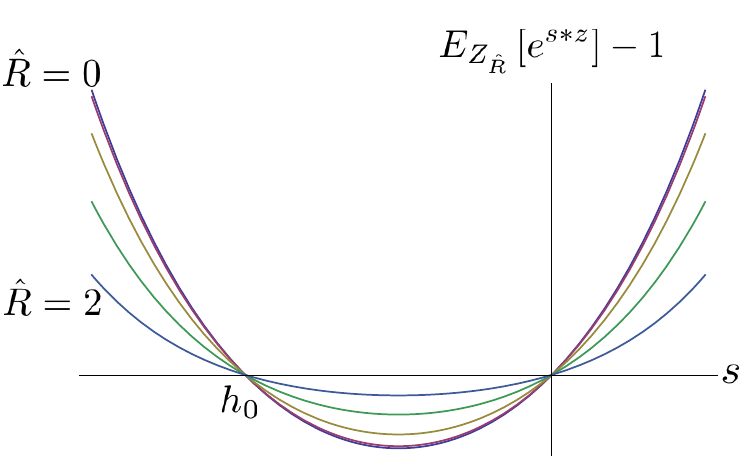}
	\end{minipage}

	\caption{The effect of $\hat{R}$ on a discrete time increment distribution, and the second real root of the moment generating function of this distribution. (A) The PDF of the random variable $Z_{\hat{R}}$, with probability mass for values between the robustness limit $\hat{R}$ re-allocated as a delta function centered at zero ($\hat{R}=1$).    (B)  The second real root $h_0$ of $M_{Z_{\hat{R}}}(s)$ remains unchanged as $\hat{R}$ increases from $0\rightarrow2$. (Lines are uniformly distributed in this range.)  This implies that in the reaction time task, no changes in the accuracy and chronometric functions will be observed until the  deviation in $E[Z_{R}]$ becomes large (Discussed in "Reaction time task: Continuous analysis").
	}
	\label{fig:Chop_PDF_H0}
\end{figure}
%FIG*FIG*FIG*FIG*FIG*FIG*FIG*FIG*FIG*FIG*FIG*FIG*FIG*FIG*FIG*FIG*FIG*FIG*FIG*FIG

	Next, we change the distribution of the accumulated increments $Z_i$ to construct a discrete time analog of the robust integrator. Specifically, increasing the robustness parameter to $R>0$ affects increments $Z_i$ by redefining the PDF $f_{Z}(Z)$ so that weak samples do not add to the total accumulated ``evidence", precisely as in Equation~\eqref{eq:chopModel}. (Models where such a central ``region of uncertainty" of the sampling distribution is ignored have previously been studied in a race-to-bound model~\cite{Smith:1989uv}; see Discussion).  This requires reallocating probability mass below the robustness limit to zero.  We plot the resulting PDF in Figure~\ref{fig:Chop_PDF_H0}(A), where the reallocated mass gives a weighted delta function at zero.   Specifically:
\begin{equation}
   f_{Z_{R}}\left(Z\right) = \delta\left(0\right) \int_{-R}^{R}f_Z\left(Z'\right)dZ' + \left\{
     \begin{array}{ccr}
       0 &:& \left|Z\right| < R \\
       f_{Z}(Z) &:& \rm{otherwise}
     \end{array}
   \right.
   \label{eq:chopPDF}
\end{equation}

%FIG*FIG*FIG*FIG*FIG*FIG*FIG*FIG*FIG*FIG*FIG*FIG*FIG*FIG*FIG*FIG*FIG*FIG*FIG*FIG
\begin{figure}
	\centering
	\includegraphics[width=6in]{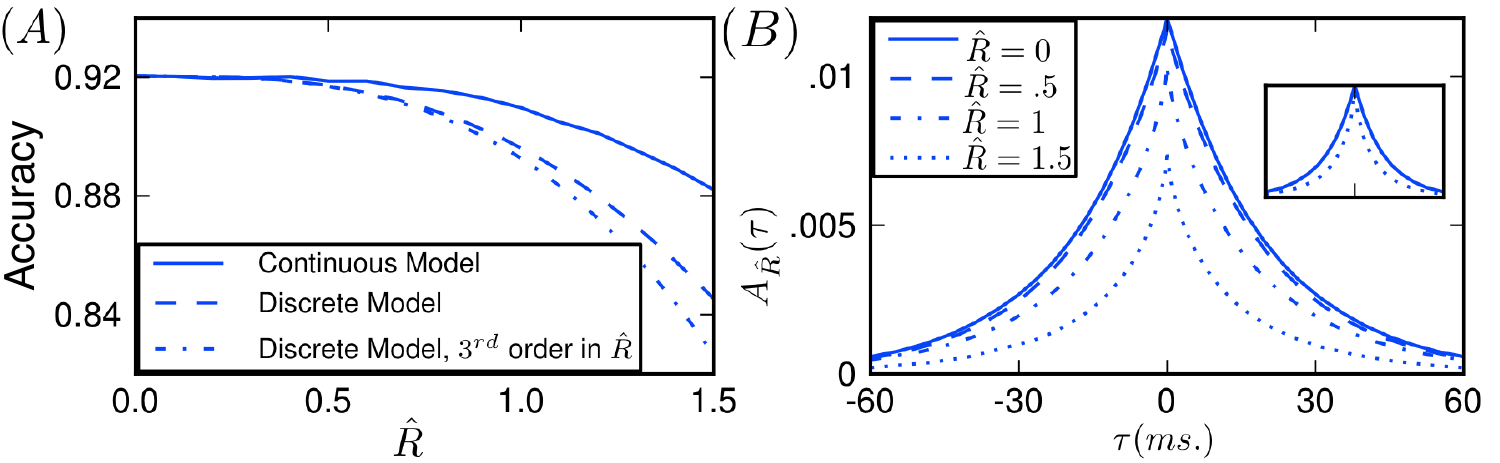}
	\caption{Comparison of the decision accuracy predicted by discrete and continuous time approximate models for the controlled duration task (as above, coherence $C=12.8$). (A) Both discrete and continuous time models predict that accuracy $(Acc$) will stay roughly constant as $\hat{R}$ increases up to $\approx 0.5$, followed by a gradual decrease.  However, the continuous time approximation provides a closer match to results for the full model pictured in Figure~\ref{fig:BasicToRobust}(A) (see text).  Additionally, comparing the dashed and dot-dashed lines shows that the approximation in Equation \ref{eq:asymptotic} provides a good description of the discrete time model for $\hat{R} \lesssim 1$. (B) The (numerically computed) autocovariance functions of the input signal $Z_{\hat R}(t)$ at various levels of $\hat{R}$ that were used to construct the continuous time curve in (A) (Equation \eqref{eq:asymptotic}). (Inset) Two of these same autocovariance functions (for $\hat{R}=0$ and $\hat{R}=1.5$) are plotted normalized to their peak value.  This shows that autocorrelation falls off faster as $\hat{R}$ increases.}
	\label{fig:FDPrediction}
\end{figure}
%FIG*FIG*FIG*FIG*FIG*FIG*FIG*FIG*FIG*FIG*FIG*FIG*FIG*FIG*FIG*FIG*FIG*FIG*FIG*FIG
	
	 The central limit theorem then allows us to approximate the new cumulative sum $E_{N_{\hat{R}}}$ as a normal distribution (for sufficiently large $N$), with $\mu$ and $\sigma$ in Equation \eqref{eq:FDPsycho} replaced by the mean and standard deviation of the PDF defined by Equation \eqref{eq:chopPDF}.  As before, we normalize $R$ by the standard deviation of the increment, $\hat{R}=\frac{R}{\sigma}$, and then express the fraction correct $Acc_{\hat{R}}$ as a function of $\hat{R}$ and $s$.  One can think of $\hat{R}$ as perturbing the original $Acc$ function (Equation ~\eqref{eq:FDPsycho}), and although this perturbation has a complicated form, we can understand its behavior by observing that its Taylor expansion has only one nonzero term up to fifth order in $\hat{R}$:
		\begin{equation}
		Acc_{\hat{R}}(N) \approx Acc(N)(1 - P(N))
		\label{eq:asymptotic}
	\end{equation}
	\begin{equation*}
		P(N) =	\frac{\sqrt{N}s\left(1+2s^{2}\right)e^{-\frac{(1+N)s^{2}}{2}}}{6\pi}\hat{R}^{3} + O(\hat{R}^5)
	\end{equation*}
Thus, for small values of $\hat R$ (giving very small $\hat R ^3$), there will be little impact on accuracy.

Equation~\eqref{eq:asymptotic} can therefore partially explain the key observation in Fig~\ref{fig:BasicToRobust}(A) that $\hat{R}$ can be increased to $0.85$ while incurring very little performance loss.
For concreteness, we focus on decisions at $T=500$ ms.  To make a rough comparison, we first assume that a new sample of evidence arrives in the discrete time model every 10 ms.  We then set the SNR $s$ so that $Acc=0.92$ for the discrete model when $N=50$, agreeing with the accuracy obtained from the continuous model at 500 ms (Fig~\ref{fig:BasicToRobust}(A)) when $R=0$.
%		\Ncomment{Deleted the reference to SNR here, was redundant}
		We then increase the robustness limit $R$.   Figure~\ref{fig:FDPrediction}(A) shows that accuracy for both the discrete time model itself (dashed line), and its approximation up to $O(\hat{R}^5)$ (dot-dashed), barely decrease at all while $\hat{R}$ is less than $\approx 0.5$, and then begin to fall off.  This is consistent with the results for the full model in Fig~\ref{fig:BasicToRobust}(A).  However, the discrete time model does predict a small decrease in accuracy at $\hat{R}=0.85$ that is not seen in the full model.  In the next section, we explain how this discrepancy can be resolved.

% - - - - - - - - - - - - - - - - - - - - - - - - - - - - - - - - - - - - - - -
	\subsubsection{Controlled duration task: Continuous time analysis}
	
	We next extend the analysis of the controlled duration task in the previous section to signal integration in continuous time.  In brief, we follow a method developed in \citetext{Gillespie:1996ve} to describe the evolution of the mean and variance of a continuous input signal that has been integrated over time.  This is challenging and interesting because, as for the signals used in modeling the random dots task above (see Methods, Sensory Input), this input signal contains temporal correlations.   As in the previous section, we describe the distribution of the integrated signal at the final time $T$, which determines accuracy in the controlled duration task.
		
	We first replace the discrete input samples $Z_i$ from the previous section with a continuous signal $Z(t)$, which we take to be a (OU) gaussian process with a correlation timescale derived from our model sensory neurons (see Methods).  We define the integrated process $$\frac{dE}{dt} = Z(t) \; \; \rightarrow \; \; E(t) = \int_0^t Z(t')dt'$$ with initial condition $E(0)=0$.  %Thus $E(t+dt)=E(t)+Z(t)dt$ for some infinitesimal $dt$.
	
	Assuming that $Z(t)$ satisfies certain technical conditions that are easily verified for the OU process (wide-sense stationarity, $\alpha$-stability, and continuity of sample paths~\cite{Gardiner:BIYQWdX3,Billingsley:Ljnv-tFk,Gillespie:1996ve}),  we can construct differential equations for the first and second moments $\left<E(t)\right>$ and $\left<E^2(t)\right>$ evolving in time. We start by taking averages on both sides of our definition of $E(t)$, and, noting that $E(0)=0$, compute the time-varying mean:
	\begin{equation}
		\frac{d\left<E(t)\right>}{dt} = \left<Z(t)\right> \implies \left<E(t)\right>=t\left<Z(t)\right> \; \;. \label{e.Emean}
	\end{equation}
	Similarly, we can derive a differential equation for the second  moment of $E(t)$:

	\begin{equation*}
		\frac{d\left<E^2(t)\right>}{dt}=2\left<Z(t)E(t)\right> \; \;.
	\end{equation*}
		The righthand side of this equation can be related to the area under the autocovariance function $A\left(\tau\right) \equiv \left<Z(t) Z(t+\tau) \right> - \left<Z(t)\right>^2$ of the process $Z(t)$:
		\begin{eqnarray*}
		\left<Z(t)E(t)\right> = \left<Z(t)\int_0^tZ(s)ds\right> &=& \int_0^t\left<Z(t)Z(s)\right>ds \\
		&=& \int_0^t\left<Z(t)Z(t-\tau)\right>d\tau \\
		&=& \int_0^t A\left(\tau\right) + \left<Z(t)\right>^2 d\tau
	\end{eqnarray*}
		We now have an expression for how the second moment evolves in time.  We can simplify the result via integration by parts:
		\begin{eqnarray}
		\left<E^2(t)\right> &=& 2 \int_0^t \int_0^s A\left(\tau\right) + \left<Z(t)\right>^2 d\tau ds = 2 \int_0^t (t-\tau) A\left(\tau\right) d\tau + t^2\left<Z(t)\right>^2 \nonumber \\
		&\implies& {\rm Var}[E(t)] = 2 \int_0^t (t-\tau) A\left(\tau\right) d\tau.
		\label{e.Evar}
	\end{eqnarray}
Because $E(t)$ is an accumulation of gaussian random samples $Z(t)$, it will also be normally distributed, and hence fully described by the mean (Equation \eqref{e.Emean}) and variance (Equation \eqref{e.Evar})~\cite{Billingsley:Ljnv-tFk}.
%identically-distributed (though not independent) samples from the $\alpha$-stable process $Z(t)$, eventually $E(t)$ will be normally distributed \cite{Billingsley:Ljnv-tFk}.  Thus, $E(t)$ is fully described by the mean (Equation \eqref{e.Emean}) and variance (Equation \eqref{e.Evar}).
	
	To model a non-robust integrator, as discussed above we take $Z(t)$ to be a OU process with steady-state mean and variance $\mu$ and $\sigma^2$, and time constant $\tau$.  For the robust case, we can follow Equation~\ref{eq:chopModel} and parameterize a family of processes $Z_{\hat{R}}(t)$ with momentary values below the robustness limit $\Rhat$ set to zero. (Here, we again normalize the robustness limit by the standard deviation of the OU process.) We numerically compute the autocovariance functions $A_{\hat{R}}\left(\tau\right)$ of these processes, and use the result to compute the required mean and variance, and hence time-dependent signal-to-noise ratio SNR(t), for the integrated process $E(t)$.   This yields
\begin{equation}
	SNR_{\hat{R}}(t) =\frac{\left<E(t)\right>}{\sqrt{Var[E(t)]}}  = \frac{tE[Z_{\hat{R}}(t)]}{\sqrt{2 \int_0^t (t-\tau) A_{\hat{R}}\left(\tau\right) d\tau}} \; \;.
	\label{eq:snr}
\end{equation}
Under the assumption that $E(T)$ is approximately gaussian for sufficiently long $T$ (which can be verified numerically), we use this SNR to compute decision accuracy at $T$:
\begin{equation}
		Acc_{\hat{R}}(T) \approx \frac{1+\text{Erf}\left(\sqrt{\frac{1}{2}}SNR_{\hat{R}}(T)\right)}{2} \; .
		\label{eq:FDContinuousTime}
\end{equation}
This function is plotted for $T=500$ ms as the solid line in Figure~\ref{fig:FDPrediction}(A).  The plot shows that accuracy remains relatively constant until the robustness limit $\hat R$ exceeds $\approx 0.85$.  Interestingly, this is a longer range of $\hat R$ values than for the discrete time case (compare dotted line in Figure~\ref{fig:FDPrediction}(A)), and is closer to the results for the full model pictured in Figure~\ref{fig:BasicToRobust}(A).

Why does the robustness limit appear to have a milder effect on degrading decision accuracy for our continuous vs. discrete time input signals?  We can get some insight into the answer by examining the autocovariance functions $A_{\hat{R}}\left(\tau\right)$, which we present in Figure~\ref{fig:FDPrediction}(B).
%The peak --- which could be viewed as a proxy for the SNR --- of the signal remains unchanged until $\hat{R} > 0.5$, and then decreases as $\hat{R}$ continues to increase.
When normalized by their peak value, the autocovariance for $\hat{R} > 0.5$ falls off more quickly vs. the time lag $\tau$ (see inset in Figure~\ref{fig:FDPrediction}(B)), indicating that subsequent samples become less correlated in time.  Thus, there are effectively more ``independent" samples that are drawn over a given time range $T$, improving the fidelity of the signal.  This effect is not present in our discrete time model.

\medskip
\noindent {\it Summary:}  Our analysis of decision performance for the controlled duration task shows that two factors contribute to the preservation of decision performance for robust integrators.
The first is that, for robustness limits up to $\hat R \approx 0.5$, the momentary SNR of the inputs is barely changed by setting values below robustness limit to zero.
The second is that, as $\hat R$ increases, the signal $Z_{\hat{R}}(t)$ being integrated becomes less correlated in time.  This means that (roughly) more independent samples will arrive over a given time period.
%Thus, even as the mean-to-variance ratio decreases with $\hat R$,

% - - - - - - - - - - - - - - - - - - - - - - - - - - - - - - - - - - - - - - -
\subsubsection{Reaction time task: Discrete analysis}
	
	 %FIG*FIG*FIG*FIG*FIG*FIG*FIG*FIG*FIG*FIG*FIG*FIG*FIG*FIG*FIG*FIG*FIG*FIG*FIG*FIG
\begin{figure}
	\centering
	\makebox[6in][l]{(A)\hspace{3in}(B)}
	\begin{minipage}[b]{6in}
		\includegraphics[width=3in]{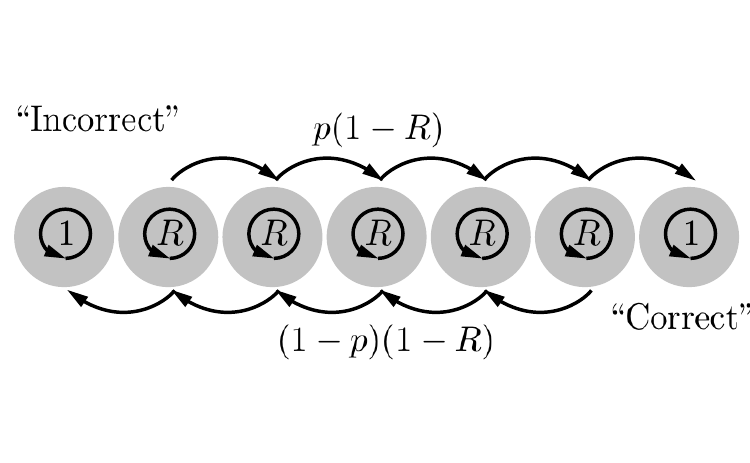}
		\includegraphics[width=3in]{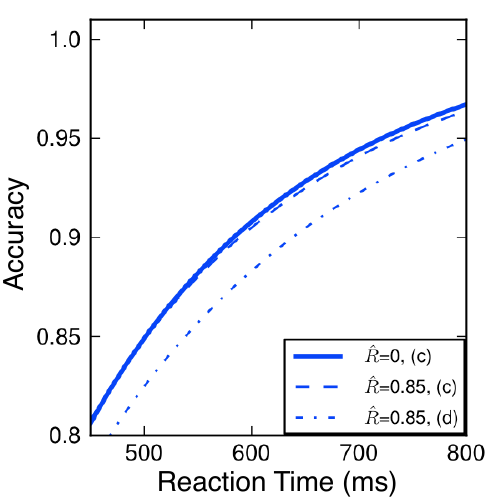}
	\end{minipage}
	\caption{Analysis of the effect of the robustness limit in the reaction time task. (A) Biased random walk between two absorbing boundaries.  The \textit{correct} and \textit{incorrect} states act as ``sinks" of the discrete time discrete space Markov chain.  The intermediate states are analogous to the potential wells in the robust integrator model.  Here, a particle will remain in the current state at the next time step with probability $R$.  The final probability of ending up in either sink is independent of $R$. (B) Speed accuracy tradeoff curves from Figure~\ref{fig:BasicToRobust} are replotted and labeled (c), while a new line labeled (d) shows the performance predicted by the discrete time, continuous space model described in "Reaction time task:  Continuous analysis".  Here the signal-to-noise ratio of the discrete increments were chosen so that the line generated in the $\hat{R}=0$ case would overlay the line from the $\hat{R}=0$ continuous model, and is therefore not plotted. We see that the discrete time model over-predicts the impact of the robustness limit $\hat{R}$, just as in the controlled duration case (Figure~\ref{fig:FDPrediction}).}
	\label{fig:markovModel}
\end{figure}
%FIG*FIG*FIG*FIG*FIG*FIG*FIG*FIG*FIG*FIG*FIG*FIG*FIG*FIG*FIG*FIG*FIG*FIG*FIG*FIG
	
	We begin our analysis of the reaction time task by introducing a discrete time, discrete space random walk model.  In this model, schematized in Figure~\ref{fig:markovModel}(A) with five intermediate states, a particle representing the accumulated value $E$ starts at a state balanced between two absorbing ``sink" states.  At every time step, the particle moves towards the ``correct" (i.e. preferred) sink with probability $p(1-R)$, and towards the ``incorrect" (null) sink with probability $(1-p)(1-R)$ (we consider $p>0.5$, biasing the random walk toward the ``correct" sink).  There is also the possibility that the particle might remain in the current state, with probability $R$.
	
	We now draw an analogy between the states in this random walk model and the ratcheting dynamics among energy wells in a robust integrator (see Figure~\ref{fig:Energy_Surface} and Introduction).  Here, the position of the particle represents the accumulated evidence for the left vs. right alternatives, and the absorbing states represent crossing of the corresponding decision thresholds.  When the robustness limit $R$ is increased, the wells --  each of which could represent a bistable neural subpopulation (see Methods) -- act to hold the particle in a given state, with a probability set by $R$.
	
	As $R$ is increased in the random walk model, the probability of transitioning out of a given state similarly decreases.  Standard results on Markov chains (see, for example, \citetext{Kemeny:VnS-Humo}) provide formulas for the probability that the particle will end in one vs. the other sink, as well as the expected number of time steps until this occurs, based on the transition matrix associated with the random walk. The probability of ending in the ``correct" sink corresponds to decision accuracy, and is found at the middle entry in the solution vector $\vect{x}$ of the matrix equation
	
	\begin{equation}
		\left(\vect{I}-\vect{Q}\right)\vect{x}=(1-R)p\vect{e_1} \;.
				\label{eq:solvewcramer}
	\end{equation}
Here $\vect{Q}$ is a  tridiagonal matrix with $R$ on the main diagonal, $p(1-R)$ on the lower diagonal, and $(1-p)(1-R)$ on the upper diagonal; $\vect{e_1}$ is the canonical basis vector with $e_1^{(1)}=1$, and all other entries equal to $0$.
	After some factoring, we find a common factor of $(1-R)$ on both sides of the equation; thus the solution to $\vect{x}$ is independent of $R$. This implies that the probability of ending up in the correct state is unchanged by increasing $R$ from the non-robust case ($R=0$).  Intuitively this makes sense: if one conditions on the fact that one will leave the current state on the next time step, the probability of moving toward the correct and incorrect states are independent of $R$.
	
	The same is not true for the expected number of steps necessary to reach a sink (by analogy, the reaction time).  This is because the matrix system that yields reaction times is:
	\begin{equation}
		\left(\vect{I}-\vect{Q}\right)\vect{t}=\vect{1}.
	\end{equation}
	Here the right-hand side of this equation is the vector of all ones, and therefore no equivalent cancellation can occur. However, we do notice that the reaction time with $R\neq0$ is just a rescaling of the original reaction time with $R=0$.  Specifically, if $t_R$ is the expected number of steps required to reach an absorbing state, then
		\begin{equation}
		t_R = t_0\frac{1}{1-R} \; \;.
	\end{equation}	
Thus, the only effect of the robustness limit $R$ is to delay arrival at the sinks.

\medskip
\noindent {\it Summary:} We have used a simplified random walk model to gain intuition about the effect of the robustness limit in the reaction time task, and to show that adding a robustness limit only affects decision latency, but not accuracy.  In the next section, we will derive a similar result for continuous sample distributions.

%	\begin{equation}
%		P[X_i = x] = \left\{
%		\begin{array}{ll}
%			p(1-R)  & : x = 1 \\
%			R  & : x = 0 \\
%			(1-p)(1-R)  & : x = -1
%		\end{array}
%		\right.
%	\end{equation}
%
%	We can give this markov chain the notation of a stochastic process by defining $X_i$ as an incremental displacement of the particle, and $Y_i$ as the number of states that the particle has moved from the initial condition after $i$ time steps. Here $Y_i$ is a signed quantity, with positive sign indicating displacement in the direction of the ``correct" alternative. $Y_i$ is itself a stochastic process indexed by $i$, such that:
%	\begin{equation}
%		Y_n = \sum_{i=1}^n X_i
%	\end{equation}
%	Wald~\cite{Wald:1944th} provides a description of how to compute the probability that the particle will arrive in the ``correct" state in terms of the roots of the moment generating function (MGF) of $X_i$, set to unity.  The essence of the technique can be interpreted as solving Equation~\ref{eq:solvewcramer} via Cramer's rule.
%	\begin{eqnarray}
%		1 &=& E[e^{tX_i}] = p(1-R)e^{(-1)t} + Re^{0t} + (1-p)(1-R)e^{t}\\
%		&\iff& 0=p(e^{-t}+(1-p)e^{t}
%	\end{eqnarray}
%	As can be seen, the roots of this expression in $t$ do not depend on $R$, the parameter that causes the particle to remain in it's current state.  In this way, we see that the accuracy of this decision making model is unaffected by robustness precisely because the roots of the MGF of the increment distribution remain fixed as $R$ increases.

% - - - - - - - - - - - - - - - - - - - - - - - - - - - - - - - - - - - - - - -
\subsubsection{Reaction time task:  Continuous analysis}  \label{s.rtcts}
	
We return to the continuous sampling distribution introduced in ``Controlled duration task: Discrete time analysis", but now in the context of threshold crossing in the reaction time task. The accumulation of these increments toward decision thresholds can be understood as the sequential probability ratio test, where the log-odds for each alternative are summed until a predefined threshold is reached \cite{Wald:1945tg,Gold:2002th,Luce:1963uz,Laming:1968vv}. \citetext{Wald:1944th} provides an elegant method of computing decision accuracy and speed (RT).  The key quantity is given by the moment generating function (MGF, denoted $M_Z(s)$ and defined in Equation \ref{eq:MGF}) for the samples $Z$ (see \citetext{Luce:1986vp} and \citetext{Doya:2007wx}, Chapter 10).  Under the assumption that thresholds are crossed with minimal overshoot, we have the following expressions:
\begin{eqnarray}
	{Acc} &=& \frac{1}{1+e^{\theta h_0}}  \label{eq:FC} \\
	\text{RT} &=& \frac{\theta}{E[Z]} \tanh\left[-\frac{\theta}{2} h_0 \right]\label{eq:RT}
\end{eqnarray}
where $h_0$ is one of the two real roots of the equation $M_Z(s)=1$ (the other root is precisely $1$) and $\theta$ is the decision threshold.

We first consider the case of a non-robust integrator, for which the samples $Z$ are again normally distributed.  In this case, we must solve the following equation to find $s=h_0$:
\begin{equation}
M_Z(s)=		E_{Z}\left[e^{s*z} \right]		= \int_{-\infty}^{\infty}f_Z\left(z\right)e^{s*z}dz     = e^{\frac{s^2 \sigma ^2}{2}+s \mu } = 1.
	\label{eq:MGF}
\end{equation}
It follows that $s=1$ and $s=h_0=-2\frac{\mu}{\sigma^{2}}$ provide two real solutions of this equation. (Wald's Lemma ensures that there are exactly two such real roots, for any sampling distribution meeting easily satisfied technical criteria.)

When the robustness limit $R > 0$, we can again compute the two real roots of the associated MGF. Here, we use the increment distribution $f_{Z_R}(Z)$ given by Equation~\eqref{eq:chopPDF}, for which all probability mass within $R$ of 0 is reassigned precisely to 0.  Surprisingly, upon plugging this distribution into the expression $M_Z(s)=1$, we find that
% Where $H(z)$ indicates the Heaviside function:
%
%\begin{eqnarray}
%	1		&=&		 \int_{-\infty}^{\infty}f_{Z_{R}}\left(z\right)e^{t*z}dz     \\
%		f_{Z_{R}}\left(z\right)	&=&  		\delta (z) \int_{-R}^{R}f_Z\left(z\right)dz + (H (z-R )-H (z+R )+1) f_Z\left(z\right) \label{eq:chopDist}
%\end{eqnarray}
%These integrals are readily calculable, and once they are computed it is easy to demonstrate that
$s=1,h_0$ continue to provide the two real solutions to this equation \emph{regardless of} $R$, as depicted in Figure~\ref{fig:Chop_PDF_H0}(B).

This observation implies that (1) accuracies (Equation \eqref{eq:FC}) are unchanged as $R$ is increased, and (2) reaction times (Equation \eqref{eq:RT}) only change when $E[{Z_{R}}]$ changes.  In other words, the integrator can ignore inputs below an arbitrary robustness limit {at no cost to accuracy}, and a penalty in terms of reaction time will only be observed when $E[{Z_{R}}]$ changes appreciably.  Generalizing our result, we note that a sufficient condition for $h_0$ to be unchanged as $R$ changes is that the original sampling distribution $f_{Z}(z)$ obeys
\begin{equation}
	f_{Z}(z)=f_{Z}(-z)e^{h_0 z};
\end{equation}
it is straightforward to verify that the Gaussian satisfies this property.

We next determine the magnitude of $R$ necessary to change $E[Z_{R}]$.  When we substitute $\hat{R}=\frac{R}{\sigma}$ and compute the perturbation to $E[Z_{R}]$, we again find only one term up to fifth order in $\hat{R}$:
\begin{equation}
	E[Z_{R}] =	\int_{-\infty}^{\infty}z \; f_{Z_{R}}\left(z\right) \; dz  = E[Z]\left(1-P\right)
	\label{eq:firstTerm}
\end{equation}
\begin{equation*}
	P =	\sqrt{\frac{2}{9\pi}} e^{-\frac{1}{2}\left(\frac{\mu}{\sigma}\right)^{2}} \hat{R}^{3} + O(\hat{R}^{5})
\end{equation*}
This outcome is similar to the controlled duration case:  small values of $\hat R$ will have little effect on $E[Z_{R}]$, and therefore little effect on increasing decision speed (via Equation ~\eqref{eq:RT}).  Moreover, as we have already shown, accuracy is unaffected by robustness limits $\hat R$ of {\it any} value.  As a consequence we expect speed accuracy curves to change only modestly for small values of $\hat R$. We illustrate this via a speed accuracy plot in Figure~\ref{fig:markovModel}(B).  Here, the present discrete time, continuous space model produces the chain-dotted curve (marked (d)), showing a moderate decrease in performance at $\hat{R}=0.85$.  This decrease is purely due to the increase in RT just discussed.

%for the  We therefore are not surprised that the speed-accuracy tradeoff curve pictured in s unaffected by setting $\hat{R}=.85$. The speed-accuracy tradeoff frontier defined by Equations \ref{eq:RT} and \ref{eq:FC} can be parameterized by the SNR, as can the perturbation in equation \ref{eq:firstTerm}. Similar to the fixed-duration case, then, this single parameter can be fit to the performance at $\hat{R}=0$, and the quality of the discrete time approximation to the results of the Monte-Carlo can be examined; this is done in Figure \ref{fig:markovModel}.

However, the model at hand does not reproduce the speed accuracy curve for the continuous time model shown in Figure~\ref{fig:markovModel}(B).  Indeed, the continuous time model produces better performance (higher accuracy at a given speed).  This suggests an additional effect in the continuous time case:  once again, the fact that $\hat R$ reduces autocorrelation of the integrated signal increases the fidelity of the input, improving performance (see inset in Figure~\ref{fig:FDPrediction}(B)).  Unlike the simpler controlled duration task, attempting a mathematical analysis of this effect is beyond the scope of this paper.

% for the first-passage density distribution over a single absorbing barrier~\cite{Hesse:1991p447}.  Because of this, our quantitative analysis cannot be extended in the same way as the FD case.  However, qualitatively we can begin to see why our initial hypotheses about the large impact of $\hat{R}$ begin to break down because of the way robustness both preserves the roots of the MGF of the sampling distribution, and begins to diminish the autocorrelation of the evidence stream.

\medskip
\noindent {\it Summary of analysis:}  We pause to summarize our analysis of how the robustness limit $R$ impacts decision performance.  For both the controlled duration and reaction time tasks, we first studied the effect of this limit on the evidence carried by momentary values of sensory inputs.  In each task, this effect was more favorable than might have been expected:  in the controlled duration case, the signal to noise ratio of momentary inputs was preserved for a fairly broad range of $R$, while in the reaction time task, $R$ was shown to affect accuracy but not speed at fixed decision threshold.  Moreover, the robustness mechanism serves to decorrelate input signals in time.  This contributes further to decision performance being preserved as the robustness limit increases.

\subsection{Reward rate and the robustness-sensitivity tradeoff}
\label{sec:fundamentalTradeoff}

%FIG*FIG*FIG*FIG*FIG*FIG*FIG*FIG*FIG*FIG*FIG*FIG*FIG*FIG*FIG*FIG*FIG*FIG*FIG*FIG
\begin{figure}[t!]
	\centering
	\includegraphics[width=6in]{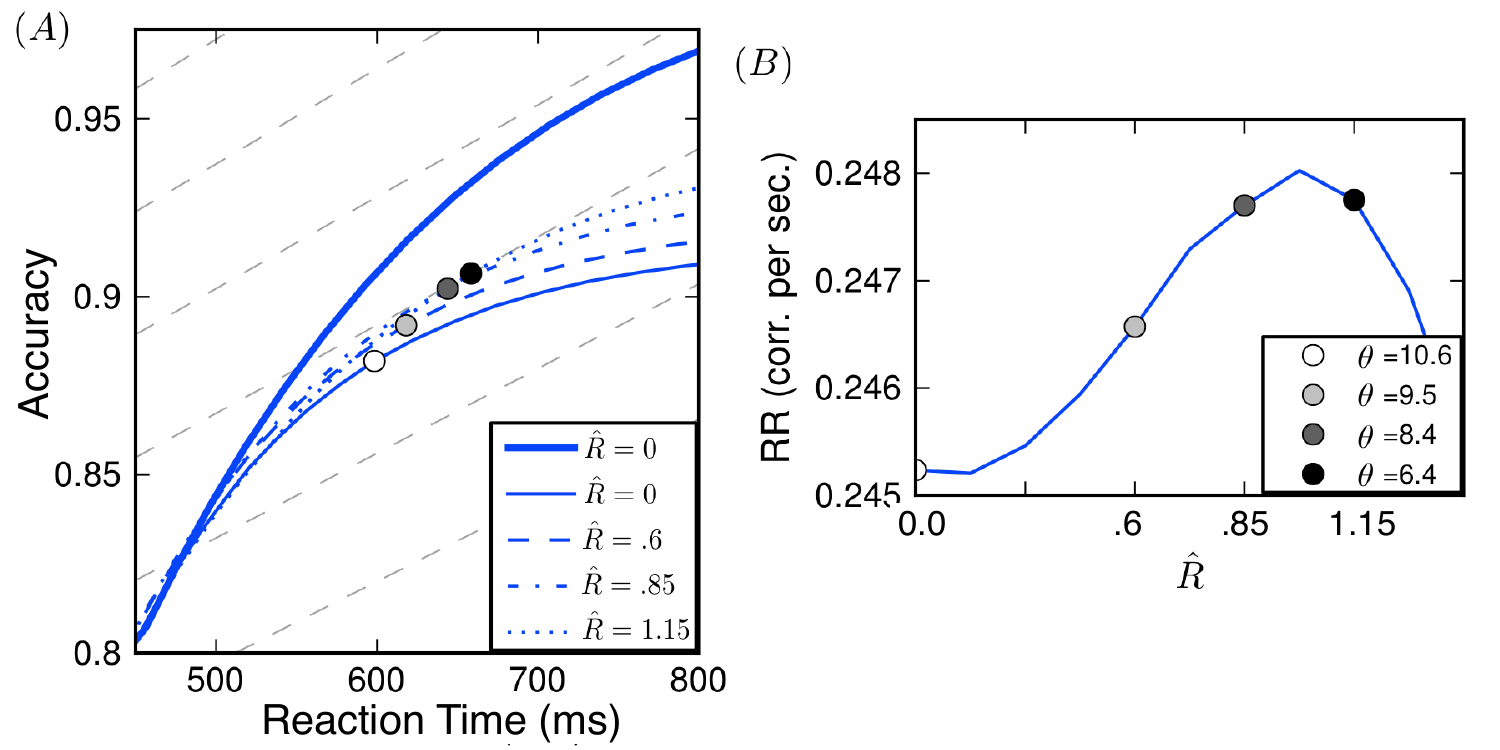}
	\caption{Using the Reward Rate metric to quantify recovery of decision performance as the robustness limit is increased, $C=12.8$.  (A) Speed accuracy curves plotted for multiple values of $\hat{R}$; as in previous figures, the greater accuracies found at fixed reaction times indicate that performance improves as $\hat{R}$ increases.  The heavy line indicates the ``baseline" case of a perfectly tuned, non-robust integrator (repeated from Figure~\ref{fig:BasicToMistuned}(B)).    RR isoclines are plotted in background (dotted lines; see text), and points along speed accuracy curves that maximize RR are shown as circles.  These maximal values of reward rate are plotted in (B), demonstrating the non-monotonic relationship between $\Rhat$ and the best achievable RR.}
	\label{fig:Effect_Of_Chop}
\end{figure}
%FIG*FIG*FIG*FIG*FIG*FIG*FIG*FIG*FIG*FIG*FIG*FIG*FIG*FIG*FIG*FIG*FIG*FIG*FIG*FIG

Until now, we have examined performance in the reaction time task by plotting the full range of attainable speed and accuracy values.  The advantage of this approach is that it demonstrates decision performance in a general way.  An alternative, more compact approach, is to assume a specific method of combining speed and accuracy into a single performance metric. This approach is useful in quantifying decision performance, and rapidly comparing a wide range of models.

%  However, the latter is difficult to quantify the degree to which a change to $\sigma_{\beta}$ or $\hat{R}$ either improves or diminishes performance across multiple model parameter choices.
%We now combine speed and accuracy into a single scalar performance metric,

Specifically, we use the reward rate (RR)~\cite{Gold:2002th,Bogacz:2006fj}:
\begin{equation}
	RR = \frac{FC}{RT+T_{del}} \; .
	\label{eq:RR}
\end{equation}
Reward rate can be thought of as the number of correct responses made per unit time, with a delay $T_{del}$ imposed between responses to penalize rapid guessing. Implicitly, this assumes a motivation on the part of the subject which may not be true; in general, subjects rarely achieve optimality under this definition as they tend to favor accuracy over speed in two-alternative forced choice trials~\cite{Zacksenhouse:2010ge}. Here, we simply use this quantity to formulate a scalar performance metric that provides a clear, compact interpretation of reaction time data.

Plotted in Figure~\ref{fig:Effect_Of_Chop}(A) are multiple accuracy vs. speed curves. The heavy solid line corresponds to the ``baseline" model with robustness and mistuning set to zero (see Figure~\ref{fig:2ParamOverview}).  The lighter solid line corresponds to the ``mistuned" model with $\sigma_{\beta}=.1$.  The remaining dashed lines correspond to the ``recovery" model for three different, nonzero levels of the robustness limit $\hat{R}$.  Also plotted in the background as dashed lines are RR isoclines -- that is, lines along which RR takes a constant value, with $T_{del}=3$ sec.  On each accuracy vs. speed curve, there exists a RR-maximizing (RT, accuracy) pair.  This corresponds to a tangency with one RR isocline, and is plotted as a filled circle.  In general, each model achieves maximal RR via a different  threshold $\theta$; values are specified in the legend of Panel (B).  (A general treatment of RR-maximizing thresholds for drift-diffusion models is given in \citetext{Bogacz:2006fj}.)

In sum, we see that mistuned integrators with a range of increasing robustness limits $\hat R$ achieve greater RR, as long as their thresholds are adjusted in concert. The optimal values of RR for a range of robustness limits $\hat R$ are plotted in Figure~\ref{fig:Effect_Of_Chop}(B).  This figure illustrates the fundamental tradeoff between robustness and sensitivity discussed above.  If there is variability in feedback mistuning ($\sigma_{\beta}>0$), increasing $\hat{R}$ can help recover performance.  However, beyond at a certain point increasing $\hat{R}$ further starts to diminish performance, as too much of the input signal is ignored.

	% - - - - - - - - - - - - - - - - - - - - - - - - - - - - - - - - - - - - - -
	\subsection{Biased mistuning towards leak or excitation}
	%FIG*FIG*FIG*FIG*FIG*FIG*FIG*FIG*FIG*FIG*FIG*FIG*FIIG*FIG*FIG*FIG*FIG*FIG*FIG
	\begin{figure}
		\centering
		\makebox[6in][l]{(A)\hspace{3in}(B)}
		\includegraphics[width=6in]{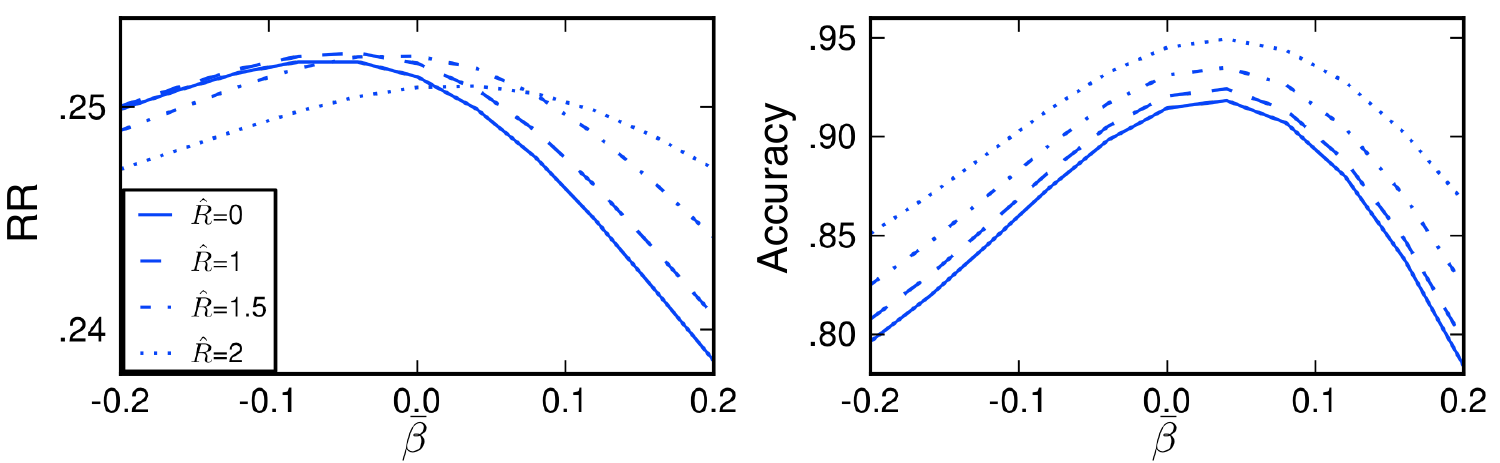}
		\caption{A nonzero robustness limit improves performance across a range of mistuning baises $\bar{\beta}$. In both the reaction time (A) and controlled duration (B) tasks, robustness helps improve performance when $\beta \sim \left(\bar{\beta},.1^2\right)$, for all values of $\bar{\beta}$ shown.  Here, as in previous panels, the coherence of the sensory input is $C=12.8$.  In the reaction time task, $\theta$ is varied for each value of $\bar{\beta}$ to find the maximal possible reward rate RR (see text);  $T_{del}=3$ seconds.  In the controlled duration case, $\theta=15.3$ is fixed, in agreement with a value indicated by behavioral data (see below and Figure~\ref{fig:naivePsychoCrono}).}
		\label{fig:BetaHat}
	\end{figure}
	%FIG*FIG*FIG*FIG*FIG*FIG*FIG*FIG*FIG*FIG*FIG*FIG*FIIG*FIG*FIG*FIG*FIG*FIG*FIG
	
	We next consider the possibility that variation in mistuning from trial to trial could occur with a systematic bias in favor of either leak or excitation, and ask whether the robustness limit has qualitatively similar effects on decision performance as for the unbiased case studied above. Specifically, we draw the mistuning parameter $\beta$ from a gaussian distribution with standard deviation $\sigma_{\beta} = 0.1$ as above, but with various mean values $\bar{\beta}$ (see Methods).  In Figure~\ref{fig:BetaHat}(A) we show reward rates as a function of the bias $\bar{\beta}$, for several different levels of the robustness limit $\hat{R}$. At each value of $\bar{\beta}$, the highest reward rate is achieved for a value of $\hat{R}>0$; that is,
regardless of the mistuning bias, there exists a $\hat{R}>0$ that will improve performance vs. the non-robust case ($\hat{R}=0$).   We note that this improvement appears minimal for substantially negative mistuning biases, but is more noticable for the values of $\bar \beta$ that yield the highest RR.  Finally, the ordering of the curves in Figure~\ref{fig:BetaHat}(A) shows that, for many values of $\bar \beta$, this optimal robustness limit is an intermediate value less than $\hat R=2$.

While Figure~\ref{fig:BetaHat} only assesses performance via a particular performance metric (RR, $T_{del}=3$ sec.), the analysis in ``Reward rate and the robustness-sensitivity tradeoff" suggests that the result will hold for other performance metrics as well.  Moreover, Figure~\ref{fig:BetaHat}(B) demonstrates the analogous effect for the controlled duration task:  for each mistuning bias $\bar \beta$, decision accuracy increases over the range of robustness limits shown.
	% When $\hat{R}=0$, optimal performance can be achieved by positively biasing the recurrent self-excitation ($\bar{\beta}>0$).

	% - - - - - - - - - - - - - - - - - - - - - - - - - - - - - - - - - - - - - -
	\subsection{Bounded integration as a model of the fixed duration task}
	
	We have demonstrated that increasing the robustness limit $\hat{R}$ can improve performance for mistuned integrators, in both the reaction time and controlled duration tasks.  In the latter, a decision is made by examining which integrator had accumulated more evidence at the end of the time interval.  In contrast, \citetext{Kiani:2008ee} argue that decisions in the controlled duration task are actually made with a decision threshold (or bound).  That is, evidence accumulates toward a bound as in the reaction time task; if accumulated evidence crosses the bound before the end of the task duration, the subject simply waits for the opportunity to report the choice, ignoring any further evidence.
	
\begin{figure}
		\begin{minipage}[b]{6in}
			\centering
			\includegraphics[width=6in]{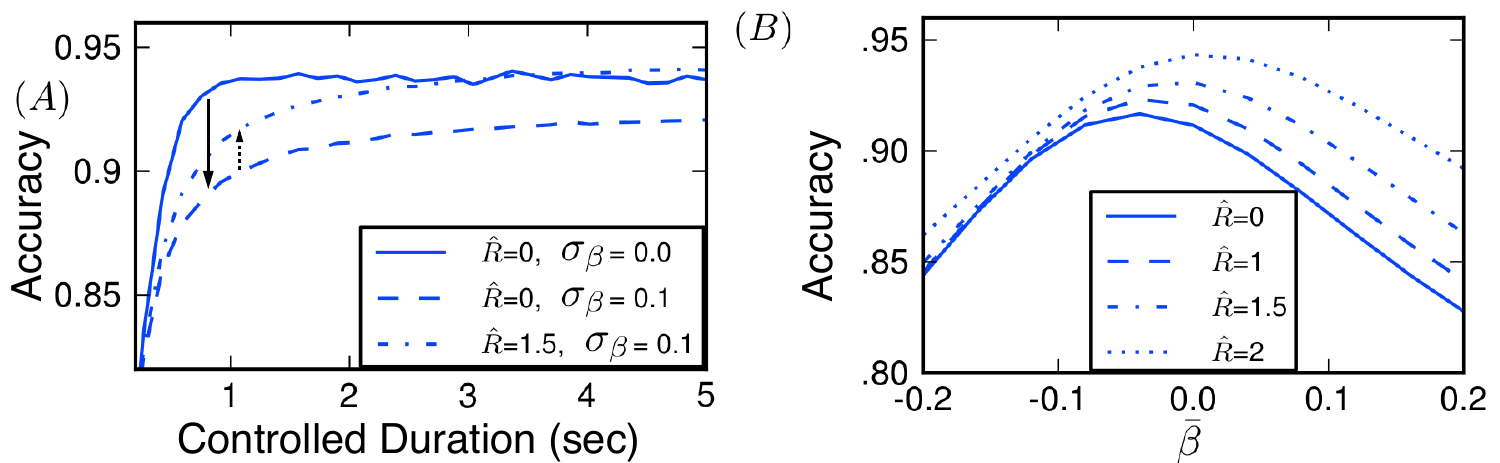}
		\end{minipage}
\caption{Effect of the robustness limit $\hat{R}$ on decision performance in a controlled duration task, under the bounded integration model of Kiani et al (2008) (see text).  Dot coherence $C=12.8$. (A) Increasing the robustness limit $\hat{R}$ helps recover performance lost to mistuning at multiple reaction times in the controlled duration task. Specifically, moving from the ``baseline" model to the ``mistuned" model decreases decision accuracy (solid arrow), but this lost accuracy can be partially or fully recovered for $\hat R>0$ (dotted arrow). (B) When allowing for biased mistuning ($\bar{\beta}\neq0$, $\sigma=.1$), $\hat{R}$ still allows for recovery of performance; effects are most pronounced when $\bar{\beta}>0$.
}
		\label{fig:FD_Beta_T_BoundNoBound}
	\end{figure}

	Figure~\ref{fig:FD_Beta_T_BoundNoBound} demonstrates that our observations about the how the robustness limit can recover performance lost to mistuned feedback carry over to this model of decision making as well.  Specifically, Panel~\ref{fig:FD_Beta_T_BoundNoBound}(A) shows how setting $\hat{R}>0$ improves performance in a mistuned integrator. In fact, more of the lost performance (up to $100\%$) is recovered than in the previous model of the controlled duration task (cf. Figure~\ref{fig:MistunedToRecovery}(A)). Panel~\ref{fig:FD_Beta_T_BoundNoBound}(B) extends this result to show that some value of $\hat{R}>0$ will recovers lost performance over a wide range of mistuning biases $\bar{\beta}$
 (cf. Figure~\ref{fig:BetaHat}(B)).
 %As before, this figure demonstrates that there is a performance recovery to be achieved through non-zero $\hat{R}$, regardless of the mean of the distribution of $\beta$.

	%FIG*FIG*FIG*FIG*FIG*FIG*FIG*FIG*FIG*FIG*FIG*FIG*FIG*FIG*FIG*FIG*FIG*FIG*FIG
	%FIG*FIG*FIG*FIG*FIG*FIG*FIG*FIG*FIG*FIG*FIG*FIG*FIG*FIG*FIG*FIG*FIG*FIG*FIG
	
	% - - - - - - - - - - - - - - - - - - - - - - - - - - - - - - - - - - - - - -
	\subsection{Compatibility of the robust integrator model with behavioral data}
\label{sec:consistent}

%FIG*FIG*FIG*FIG*FIG*FIG*FIG*FIG*FIG*FIG*FIG*FIG*FIG*FIG*FIG*FIG*FIG*FIG*FIG*FIG
\begin{figure}
	\begin{center}
	\makebox[6in][l]{(A)\hspace{3in}(B)}
	\includegraphics[width=6in]{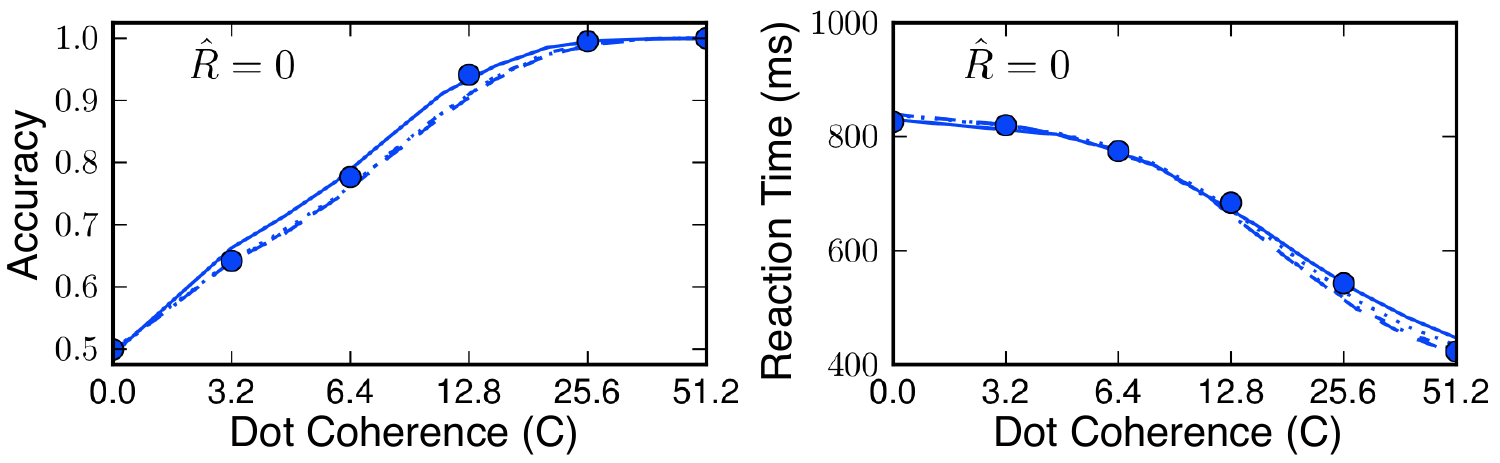}
	\makebox[6in][l]{(C)\hspace{3in}(D)}
	\includegraphics[width=6in]{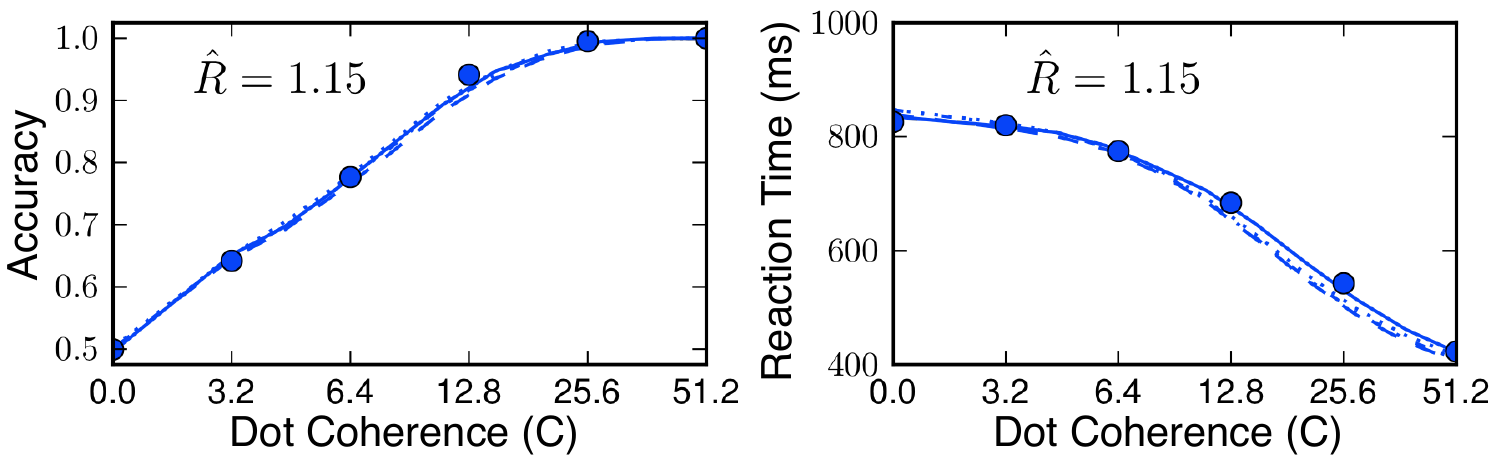}
	
		\begin{tabular}{|l||c|c|c|c|}
			\hline
							& \multicolumn{2}{c|}{} 					& Not Robust ($\hat{R}=0$) 				& Robust ($\hat{R}=1.15$)\\
			\cline{2-5}
							& \multicolumn{2}{c|}{$\beta$}					& \multicolumn{2}{c|}{{\boldmath $\left(\theta,\sqrt{\nu_{\gamma}}\right)$}}\\
			\hline \hline
			Perfect Tuning 	& $\sim \delta(0)$				& --- 			& {\boldmath $\left(15.3,14.0\right)$}	&{\boldmath $\left(7.6,14.0\right)$}\\
			\hline
			Mistuning		& $\sim N\left(0,.1^2\right)$		& - -			& {\boldmath $\left(11.0,12.0\right)$}	&{\boldmath $\left(6.8,12.0\right)$}\\
							& $\sim N\left(-.05,.1^2\right)$	& -.-			& {\boldmath $\left(9.2,12.1\right)$}		&{\boldmath $\left(5.7,12.1\right)$}\\
							& $\sim N\left(.05,.1^2\right)$	& ...			& {\boldmath $\left(13.8,11.5\right)$}	&{\boldmath $\left(8.6,11.5\right)$}\\
			\hline
		\end{tabular}

	\end{center}

	\caption{Accuracy (A,C) and chronometric (B,D) functions:  data and model predictions.   Solid dots are behavioral data for rhesus monkeys performing the dot-motion discrimination task~\protect\cite{Roitman:2002wr}.  In each panel, the accuracy and chronometric functions are identified with behavioral data via a least-squares fitting procedure over the free parameters $\theta$ and $\nu_\gamma$.  In Panels (A,B), the robustness threshold $\hat R=0$, and results are shown for ``baseline" and exemplar ``mistuned" models (see legend in table).  In Panels (C,D), $\hat R = 1.15$, and results are shown for  ``robust" and ``recovery" models.  The close matches to data points indicate that each model can be broadly reconciled with known psychophysics.  Parameter values for each curve are summarized in the included table.}
		%Psychometric and (B,D) Chronometric  functions were constructed by integrating an OU process using the robust neural integrator defined in Equation \eqref{eq:chopModel}. In the chronometric function, NDT=$350$ ms.   .  Importantly, only an adjustment of the decision bound $\theta$ aligns the model to data (A,B $\rightarrow$ C,D), despite only integrating $25\%$ of the input signal on average, and having the mistuning parameter $\beta$ vary with a standard deviation of as high as $.1$.
	\label{fig:naivePsychoCrono}
\end{figure}
%FIG*FIG*FIG*FIG*FIG*FIG*FIG*FIG*FIG*FIG*FIG*FIG*FIG*FIG*FIG*FIG*FIG*FIG*FIG*FIG

Given the fact that the robustness property can improve decision performance in our model, we next ask whether robust limits $\hat{R} > 0$ are compatible with known behavioral data.  To answer this question, we fit accuracy and chronometric functions from robust integrator models to behavioral data of \citetext{Roitman:2002wr} in the reaction time task.  This fit is via least squares across the range of coherence values, and requires two free parameters: additive noise variance $\nu_{\gamma}$ (see Methods) and the decision bound $\theta$.

Figure~\ref{fig:naivePsychoCrono} shows the results.  Panels (A) and (B) display accuracy and chronometric data (dots) together with fits for various integrator models.  First, the solid line gives the fit for the ``baseline" model.  The close match between model and data agrees with findings of prior studies~\cite{Mazurek:2003cm}.  Next, the dashed and dotted lines give fits for mistuned models ($\sigma_\beta = 0.1$), with three values of bias in feedback mistuning ($\bar \beta$).  To obtain these fits, both $\nu_{\gamma}$ and $\theta$ are changed from their values for the baseline case.  In particular, the noise variance $\nu_{\gamma}$ is lowered when feedback is mistuned.  This makes intuitive sense:  we have seen in Figure~\ref{fig:BasicToMistuned} that mistuned feedback worsens performance for a given signal, so that matching a fixed dataset with a mistuned integrator requires improving the fidelity of the incoming signal.

Figures~\ref{fig:naivePsychoCrono}(C),~(D) show analogous results for robust integrators.  For all cases in these panels, we take the robustness limit $\Rhat=1.15$.  We {\it fix} levels of additive noise to values found for the non-robust case above, on order to demonstrate that by adjusting the decision threshold, one can obtain approximate fits to the same data.  This is expected from our results above:  Figure~\ref{fig:MistunedToRecovery} shows that, while accuracies at given reaction times are higher for mistuned robust vs. non-robust models, the effect is modest on the scale of the full range of values traced over an accuracy curve.  Moreover, for the perfectly tuned case, accuracies at given reaction times are very similar for robust and non-robust integrators (Figure~\ref{fig:BasicToRobust}, with a slightly lower value of $\Rhat$).  Thus, comparable pairs of accuracy and RT values are achieved for robust and non-robust models, leading to similar matches with data.  In sum, the accuracy and chronometric functions in Figure~\ref{fig:naivePsychoCrono} show that all of the models schematized in Figure~\ref{fig:2ParamOverview} ---``baseline", ``mistuned", ``robust", and ``recovery" --- are generally compatible with the chronometric and accuracy functions reported in \citetext{Roitman:2002wr}.

	%FIG*FIG*FIG*FIG*FIG*FIG*FIG*FIG*FIG*FIG*FIG*FIG*FIG*FIG*FIG*FIG*FIG*FIG*FIG
	\begin{figure}
		\includegraphics[width=6in]{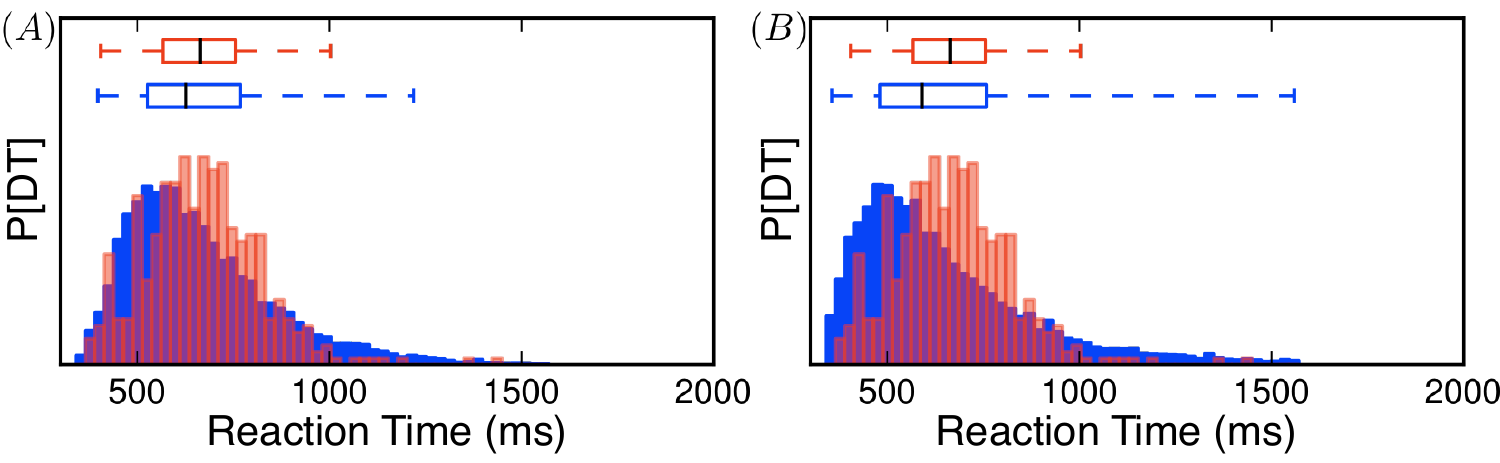}
		\caption{Reaction time histograms, with decision bounds held constant in time and dot coherence C=12.8.   Histograms for reaction times for our model with both $\hat{R}=0$ (A) and $\hat{R}=1.15$ (B) are plotted in blue.  Overlayed are the reaction time histograms for one subject (Subject ``B") in \protect\cite{Roitman:2002wr}, in red (semitransparent). Box-and-whisker plots indicate the quartiles for each data set.  Both histograms have identical means, but clearly differ in basic shape (i.e., the model produces longer tails).  We emphasize that this mismatch is a property of our basic integration to bound model, regardless of the value of the robustness limit $\hat{R}$ (see text).}
		\label{fig:RTDistCons}
	\end{figure}
	%FIG*FIG*FIG*FIG*FIG*FIG*FIG*FIG*FIG*FIG*FIG*FIG*FIG*FIG*FIG*FIG*FIG*FIG*FIG

	%FIG*FIG*FIG*FIG*FIG*FIG*FIG*FIG*FIG*FIG*FIG*FIG*FIG*FIG*FIG*FIG*FIG*FIG*FIG
	\begin{figure}
		\includegraphics[width=6in]{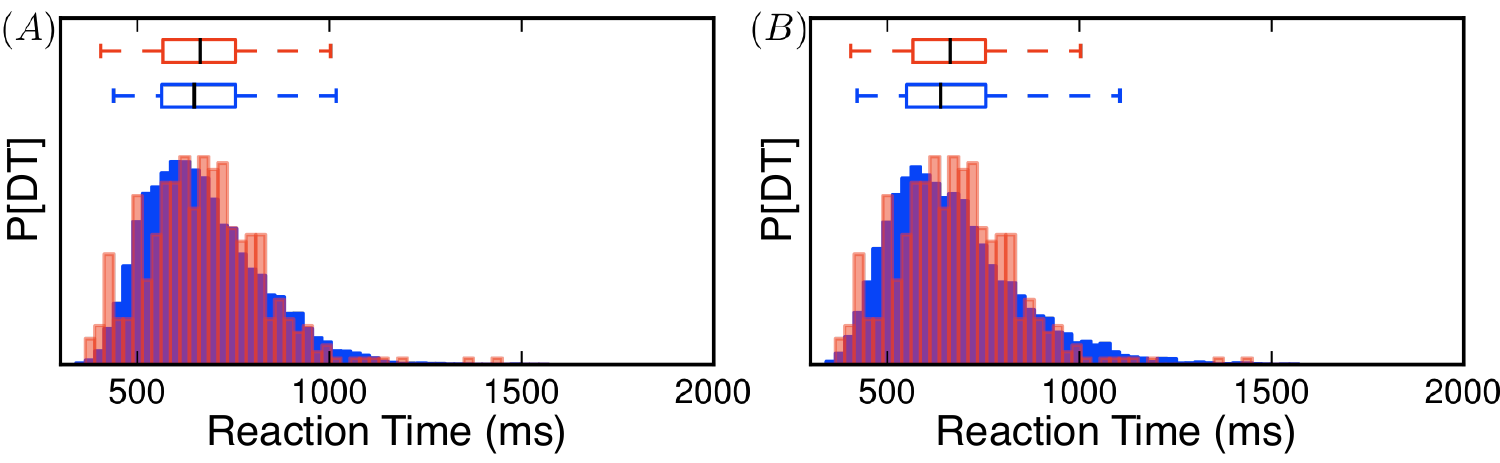}
		\caption{Reaction time histograms, with collapsing decision bounds.    This shows that introducing collapsing bounds produces simulated RT histograms that are a closer match to data, for both non-robust (Panel A, $\hat R = 0$) and robust (Panel B, $\hat R = 1.15$) integrators.  This shows that RT histograms can be similarly well matched to data for both cases. Here $t_{\frac{1}{2}}=500$, $\theta_0=25$, and $\theta_{ss}=0$.}
		\label{fig:RTDistCollapse}
	\end{figure}
	%FIG*FIG*FIG*FIG*FIG*FIG*FIG*FIG*FIG*FIG*FIG*FIG*FIG*FIG*FIG*FIG*FIG*FIG*FIG

In order to further test whether empirical data are consistent with the robust integrator model, we compared simulated reaction time histograms with those found in \citetext{Roitman:2002wr}.  First, Figure~\ref{fig:RTDistCons} compares the reaction time histograms resulting from the ``baseline" model (Panel A) and the ``recovery" model (Panel B).  These are plotted in blue;  the red histograms are data are taken from Subject ``B".  In both panels, the histograms have similar means, but differ in their shape; in particular, the model predicts a broader range of reaction times and a more slowly decaying tail of the RT distribution. From these data, we conclude that neither the ``baseline" nor the ``recovery" model quantitatively reproduce the details of reaction time distributions, when the free parameters $\theta$ and $\sigma_{\beta}$ are constrained by fitting the accuracy  and chronometric functions.

		An \textit{urgency signal} was introduced in \citetext{Churchland:2008wg} to better capture behavioral and physiological data.  We next incorporated such a signal into our model to determine whether it would better align our predicted reaction time histograms with the empirical data.  We chose to implement urgency by assuming a collapsing decision bound, which decreases monotonically from a peak value of $\theta_0$ to a steady state value $\theta_{ss}$ with a halflife $t_{1/2}$:
	\begin{equation}
		\theta(t)=\theta_0 - (\theta_0-\theta_{ss})\frac{t}{t+t_{\frac{1}{2}}}.
		\label{eq:collapsingBound}
	\end{equation}
	Figure~\ref{fig:RTDistCollapse} compares model reaction time histograms produced with the collapsing bound against the data, and indeed finds a closer fit:  qualitatively, the improvement in fit is similar for both the non-robust ($\hat R=0$) and robust ($\hat R = 1.15$) cases.  In sum, this shows that the robust integrator model is capable of producing roughly similar patterns of reaction times compared with those observed experimentally.

\section{Discussion}

A wide range of cognitive functions require the brain to process information over time scales that are at least an order of magnitude greater than values supported by membrane time constants, synaptic integration, and the like. Integration of evidence in time, as occurs in simple perceptual decisions, is one such well studied example, whereby evidence bearing on one or another alternative is gradually accumulated over time. This is formally modeled as a bounded random walk or drift-diffusion process in which the state (or decision) variable is the accumulated evidence for one choice and against the alternative(s). Such formal models explain both the speed and accuracy of a variety of decision-making tasks studied in both humans and nonhuman primates~\cite{Ratcliff:1978wz,Luce:1986vp,Gold:2007fo,Palmer:2005wa}, and neural correlates have been identified in the firing rates of neurons in the parietal and prefrontal association cortex~\cite{Mazurek:2003cm,Gold:2007fo,Churchland:2008wg,Shadlen:1996wj,Schall:2001uv,Shadlen:2001ve,Kim:2008gd}. The obvious implication is that neurons must somehow integrate evidence supplied by the visual cortex, but there is mystery as to how.
 
The reason this is a challenging problem is that the biological building blocks operate on relatively short time scales. From a broad perspective, the challenge is to assemble neural circuits that that can sustain a stable level of activity (i.e., firing rate) and yet retain the capability to increase or decrease firing rate when perturbed with new input (e.g., momentary evidence). A well known solution is to suppose that recurrent excitation might balance perfectly the decay modes of membranes and synapses~\cite{springerlink:10.1007/BF00320393,Usher:2001vq}. However, this balance must be fine tuned~\cite{Seung:1996va,Seung:2000uv}, or else the signal will either dissipate or grow exponentially (Figure \label{fig:Energy_Surface}(A), top). Several investigators have proposed biologically plausible mechanisms that mitigate somewhat the need for such fine tuning~\cite{Lisman:1998wf,Goldman:2003ge,Goldman:2009ko,Brody:2003dk,Miller:2006fp,Koulakov:2002kx}. These are important theoretical advances because they link basic neural mechanism to an important element of cognition and thus provide grist for experiment.
 
Although they differ in important details, many of the proposed mechanisms can be depicted as if operating on a scalloped energy landscape with relatively stable (low energy) values, which are robust to noise and mistuning in that they require some activation energy to move the system to a larger or smaller value (Figure \label{fig:Energy_Surface}(A), bottom; cf.~\cite{Pouget:2002va}).  The energy landscape is a convenient way to view such mechanisms -- which we refer to as robust integrators -- because it also draws attention to a potential cost. The very same effect that renders a location on the landscape stable also implies that the mechanism must ignore information in the incoming signal (i.e., evidence). Here, we have attempted to quantify the costs inherent in this loss. How much loss is tolerable before the circuit misses substantial information in the input?  How much loss is consistent with known behavior and physiology? 
 
We focused our analyses on a particular well-studied task because it offers critical benchmarks to assess both the potential costs of robustness to behavior and a gauge of the degree of robustness that might be required to mimic  neurophysiological recordings with neural network models.  Moreover, we know key statistical properties of the signal and noise to be accumulated over time, based on firing properties in area MT. 

Our central finding is that ignoring a surprisingly large part of the motion evidence would have almost negligible impact on performance. Indeed, we found that speed and accuracy are preserved even when almost a full standard deviation of the input distribution is ignored. We also found that a similar degree of robustness provides protection of performance against mistuning of recurrent excitation.  Although in general this protection is only partial (Figure \ref{fig:MistunedToRecovery}), for the controlled duration task it can be nearly complete (Figure \ref{fig:FD_Beta_T_BoundNoBound}(A), $T>2$) depending on the presence of a decision bound.

We can appreciate the impact of robust integration intuitively by considering the distribution of random values that would increment the stochastic process of integrated evidence. Instead of imagining a scalloped energy surface, we simply replace all the small perturbations in integrated evidence with zeros. Put simply, if a standard integrator would undergo a small step in the positive or negative direction, a robust integrator instead stays exactly where it was. In the setting of drift-diffusion, this is like removing a portion of the distribution of momentary evidence (the part that lies symmetrically about zero) and replacing the mass with a delta function at 0.  At first glance this appears to be a dramatic effect -- see the illustration of the distributions in Figure \ref{fig:Chop_PDF_H0} -- and it is surprising that it would not result in strong changes in accuracy or reaction time or both.
 
Three factors appear to mitigate this loss of momentary evidence. First, we showed that setting weak values of the input signal to zero can reduce both its mean and its standard deviation by a similar amount, creating compensatory effects that result in a small change to the input signal-to-noise ratio.  Second, we showed that, surprisingly, the small loss of signal to noise that does occur would not result in any loss of accuracy if the accumulation were to the same bound as for a standard integrator. The cost would be to decision time, but mainly in the regime that is dominated by drift -- that is, the shorter decision times -- hence not a large cost overall. Third, even this slowing is mitigated by the temporal dynamics of the input. Unlike for idealized drift diffusion processes, real input streams possess finite temporal correlation. Left unchecked, this would imply greater variability in the integrated signal. Interestingly, removing the weakest momentary inputs reduces the temporal correlation of the noise component of the input stream. This can be thought of as allowing more independent samples in a given time period, thereby improving accuracy at a given response time.
 
Our robust integrator framework shares features with existing models in sensory discrimination. The interval of uncertainty model of \citetext{Smith:1989uv} and the gating model of \citetext{Purcell:2010jo} ignore part of the incoming evidence stream, yet they can explain both behavioral and neural data.  We suspect that the analyses developed here might also reveal favorable properties of these models. Notably, some early theories of signal detection also featured a threshold, below which weaker inputs fail to be registered --  so called high threshold theory (reviewed in \cite{Swets:1961va}).  The primary difference in the current work is to consider single decisions made based on an accumulation of many such thresholded samples (or a continuous stream of them).  
 
Although they are presented at a general level, our analyses make testable predictions. For example, they predict that pulses of motion evidence added to random dot stimulus would affect decisions in a nonlinear fashion consistent with a soft threshold. Such pulses are known to affect decisions in a manner consistent with bounded drift diffusion ~\cite{Huk:2005gh} and its implementation in a recurrent network \cite{Wong:2007tx}. A robust integration mechanism further predicts that brief, stronger pulses will have greater impact on decision accuracy than longer, weaker pulses containing the same total evidence. 
 
However, we believe that the most exciting application of our findings will be to cases in which the strength of evidence changes over time, as expected in almost any natural setting.  One simple example is for task stimuli that have an unpredictable onset time, and whose onset is not immediately obvious.  For example, in the moving dots task, this would correspond to subtle increases in coherence from a ÒbaselineÓ of zero coherence.  Our preliminary calculations agree with intuition that robust integrator mechanism will improve performance:  in the period before the onset of coherence, less baseline noise would be accumulated; after the onset of coherence, the present results suggest that inputs will be processed with minimal loss to decision performance -- despite the continued ignoring of weak components.  This intuition can be generalized to apply to a number of settings with non-stationary sensory streams.

Many cognitive functions evolve over time scales that are much longer than the perceptual decisions we consider in this paper. Although we have focused on neural integration, it seems likely that many other neural mechanisms are also prone to drift and instability. Hence, the need for robustness may be more general. Yet, it is difficult to see how any mechanism can achieve robustness without ignoring information. If so, our finding may provide some optimism. Although we would not propose that ignorance is bliss, in the right measure it may be less costly than one would expect. 

\nocite{*}
\bibliographystyle{jneurosci}
\bibliography{robustIntegrator2}

\end{document}

\newpage

%%%%%%%%%%%%%%%%%%%%%%%%%%%%%%%%%%%%%%%%%%%%%%%%%%%%%%%%%%%
% Title Page
%%%%%%%%%%%%%%%%%%%%%%%%%%%%%%%%%%%%%%%%%%%%%%%%%%%%%%%%%%%

\noindent Title:

\noindent Supplementary Materials for Neural integrators for decision making:  A favorable tradeoff between robustness and sensitivity
\vspace{.5cm}

\noindent Abbreviated Title:

\noindent Supplementary Materials for Neural integrators for decision making
\vspace{.5cm}

\noindent Authors:

\noindent Nicholas Cain, Department of Applied Mathematics, University of Washington

\noindent Andrea K. Barreiro, Department of Applied Mathematics, University of Washington

\noindent Michael Shadlen, Department of Physiology and Biophysics, University of Washington

\noindent Eric Shea-Brown, Department of Applied Mathematics, University of Washington
\vspace{.5cm}

\noindent Corresponding Author:

\noindent Eric Shea-Brown

\noindent Department of Applied Mathematics

\noindent Box 325420

\noindent Seattle, WA 98195-2420

\noindent etsb@uw.edu

\vspace{.5cm}

\noindent Number of pages: 10

\noindent Number of figures: 7

\noindent Number of tables: 0

\vspace{.5cm}

%\noindent Contents of online supplement:
%
%\noindent In the online supplement, we include additional studies of compatibility of the robust integrator models with behavioral data from two-alternative decision making tasks.  Specifically, we compare RT histograms separated into correct and incorrect trials, and report quantitative fits with controlled duration accuracy measurements reported in \citetext{Kiani:2008ee}.  We also report ramping, trial-averaged firing rate functions that qualitatively agree with those reported in \citetext{Roitman:2002wr}, and decision-triggered stimulus traces in agreement with the data of \citetext{Kiani:2008ee}.  We also review how the simplified integrator model used in this paper can be related to the dynamics of coupled neural populations.
%\vspace{.5cm}

\noindent Conflict of Interest: None

\paragraph{Acknowledgements:}
This research was supported by a Career Award at the Scientific Interface from the Burroughs-Wellcome Fund (ESB), the Howard Hughes Medical Institute, the National Eye Institute Grant EY11378, and the National Center for Research Resources Grant RR00166 (MS), by a seed grant from the Northwest Center for Neural Engineering (ESB and MS), and by NSF Teragrid allocation TG-IBN090004.

\newpage

\section{Derivation of effective model}

Here we we describe a family of neural circuit models, parameterized by a robustness value $R$, that produce dynamics similar to that of the central robust integrator model of the main paper (given by Equation 5 in the main text).  In particular, as $R \rightarrow 0$, the dynamics reduce to a perfect integrator.  Our construction follows closely that of ~\citetext{Koulakov:2002kx} and \citetext{Goldman:2003ge}.

The circuits that we will study derive their robustness from multistability, which follows from recurrent self-excitation among multiple subpopulations (or subunits).  We begin with a system of $N$ differential equations, one for each subunit.  Depending on the activity in the rest of the network, each individual subunit can be become bistable, so that its eventual steady-state value (i.e., ``On" or ``Off")  depends on its past.  This effect,  known as hysteresis, underlies several robust integrator models~\cite{Koulakov:2002kx,Goldman:2003ge}.

The circuit integrates inputs via sequential activation of subunits, in an order determined by graded levels of ``background" inputs (or biases) to each subunit.  Following \citetext{Goldman:2003ge}, we collapse the $N$ differential equations that describe individual subunits into a equation that approximates the dynamics of the entire integrator.  This expression for the total firing rate $E(t)$ averaged over all subpopulations reduces to the robust integrator model of the main text (Equation 5 in the main text).

 % then a function of an input signal $\Delta I(t)$ and the biophysically based parameters that are defined in the subunit equations.

% - - - - - - - - - - - - - - - - - - - - - - - - - - - - - - - - - - - - - -
\subsection{Firing rate model}
The firing rate $r_i(t)$ of the $i$th bistable subunit ($i \in \{1,2,...,N \}$) is modeled by a firing rate equation:
\begin{equation}
	\tau_E \frac{dr_{i}}{dt}=-r_i+f(r_{i})
	\label{eq:fullSystem}
\end{equation}
\begin{equation*}
	f(r_{i}) = r^{-}+(r^{+}-r^{-})H\left[p*r_{i}+q(1+\beta)\sum_{i\neq j}^Nr_j+a\Delta I- b_i\right]
\end{equation*}
The parameters and variables in this equation are as follows:
	
\begin{itemize}
	\item $r_i(t)$: Firing rate of subunit (or pool)
		\item $a\Delta I(t):$ Input signal, with synaptic weight
	\item $p$: Within-pool ``synaptic" weight
	\item $q$: Between-pool ``synaptic" weight
	\item $\beta$: Fractional mistuning of feedback connectivity
	\item $N$: Number of subunits in the integrator network
	\item $r^-$: Minimum firing rate of a subunit
	\item $r^+$: Maximum firing rate of a subunit
	\item $\tau_E:$ Time constant governing firing rate dynamics
\end{itemize}
We also define:
\begin{itemize}
	\item $E(t)=\frac{1}{N} \sum_{i=1}^N r_{i}(t)$: Overall integrator activity
\end{itemize}

%FIG*FIG*FIG*FIG*FIG*FIG*FIG*FIG*FIG*FIG*FIG*FIG*FIG*FIG*FIG*FIG*FIG*FIG*FIG*FIG
\begin{figure}
	\makebox[6in][l]{(A)\hspace{3in}(B)}
	\begin{minipage}[b]{3in}
		\includegraphics[width=3in]{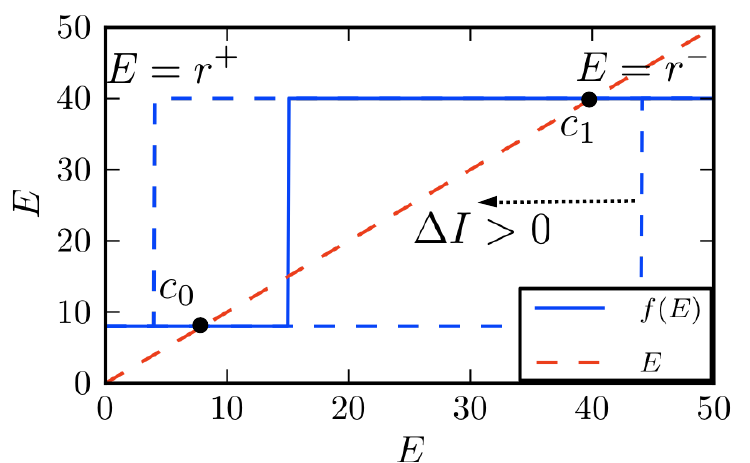}
	\end{minipage}
	\begin{minipage}[b]{3in}
		\centering
		\includegraphics[width=3in]{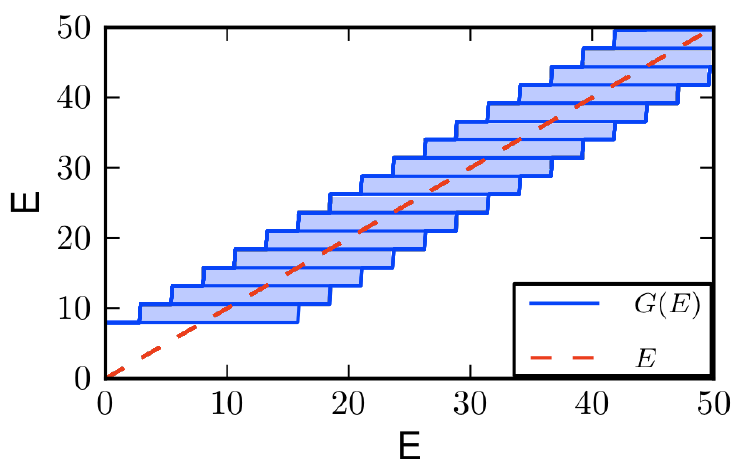}
	\end{minipage}
	\caption{Simultaneous plots of the identity line and the ``feedback line," $G(E)$ for two circuits with different numbers of subunits. (A) Here $N=1$, and so the feedback line  $G(E)=f(E)$ is exactly determined as a function of $E=r_1$ (see Equation~\eqref{eq:G}). The two intersections $c_0$ and $c_1$ are stable fixed points. In this way, the subunit's firing rate is bistable, and the value attained will depend on the history of the circuit activity.  As $\Delta I$ is changed, this translates $f(E),$ eventually eliminating either $c_o$ or $c_1$ and forcing the subunit to the remaining stable fixed point (here, $c_0$ corresponds to $E=8$ Hz. and $c_1$ to $40$ Hz).  In dashed curves, the feedback line is plotted for two such values of $\Delta I$.
%Additionally, the feedback line for the integrator in the off state ($E=r^-$) and the on state ($E=r^+$) are also plotted. 
(B) Now $N>1$ and so the feedback line $G(E)$ is no longer unambiguously specified as a function of its argument.  The function is instead the sum of $N$ potentially bi-valued functions, who's actual values will depend on the stimulus history. We represent this fact by plotting the feedback line as a set of stacked boxes, representing the potential contribution of the $i^{th}$ subunit to the total integrator dynamics~\protect\cite{Goldman:2003ge}.}
	\label{fig:SingleUnitDots_Comb}
\end{figure}

Figure \ref{fig:SingleUnitDots_Comb}(A) demonstrates the firing rate dynamics of a circuit composed of a single subunit ($N=1$).  Here we refer to Equation~\eqref{eq:fullSystem} and plot two curves vs. $E=r_1$.  The first is the identity line, corresponding to the first ``decay" term in Equation~\eqref{eq:fullSystem}.  The second is the ``feedback" line, defined by the second term in Equation~\eqref{eq:fullSystem}.  Since $N=1$, this simplifies to $f(r_1)$.   The two intersections marked $c_0$ and $c_1$ are stable fixed points (which we refer to as ``On" and ``Off" respectively).  Thus, the subunit shown is bistable. Importantly, however, the location of the step in $f(E)$ varies with changes in the input signal (as per Equation~\ref{eq:fullSystem}).  In particular, substantial values of $\Delta I(t)$ will (perhaps transiently) eliminate one of the fixed points, forcing the subunit into either the ``On" or the ``Off" state with $r_i = r_+$ or $r_i = r_-$, respectively.  Moreover, the change is self-reinforcing via the recurrent   excitation $pr_i$.  The range over which a given subunit displays bistability is affected by the mistuning parameter $\beta$, which scales the total recurrent excitation from the rest of the circuit.

%Following Goldman et al.~\cite{Goldman:2003ge}, we define the total network activity $E(t)$ via

We now derive the dynamics of the overall firing rate $E(t).$  After summing both sides of Equation \eqref{eq:fullSystem} over $i$ we have:
\begin{equation}
	\tau_E \frac{dE}{dt} = -E+G(E) 		\label{eq:sumBothSides}
\end{equation}
where
\begin{equation}
	G(E) = r^{-}+\frac{(r^+-r^-)}{N}\sum_{i=1}^NH\left[(p-q(1+\beta))r_{i}+Nq(1+\beta)E+a\Delta I- b_i\right] \; \;. \label{eq:G}
\end{equation}
At this point, we almost have a differential equation for a single variable, $E(t)$. However, Equation~\eqref{eq:G} still depends on the $N$ activities $r_i$ of the individual subunits, and at any particular time their values are not uniquely determined by the value of $E$; we can only bound their values as $r^- \le r_i \le r^+$.  We will return to discuss the dynamics of Equation~\ref{eq:sumBothSides} below.

\subsection{Bias term}

The bias term for the $i^{th}$ subunit, $b_i$, is set by analyzing the range of values of $E$ for which the exact value of the feedback function $f(r_i)$ is unknown.  In the case of an integrator composed of only a single subunit, we choose the bias term to cause the positive input needed to force the unit to be on, and the negative input needed to force the unit to be off, to take the same values.  This yields $b_1=\frac{p(r^++r^-)}{2}$.

The general case of $N$ subunits is more complicated.  Now the feedback contribution of the $i^{th}$ unit, $f(r_i)$, is no longer a simple function of the population activity $E$.  Instead, it has additional dependence on its own activity $r_i$.  We see this clearly in Equation \ref{eq:G}, where the values of $r_i$ that contribute to the definition of $G(E)$ are unspecified.  However, we do know that each $r_i$ is trapped between $r^+$ and $r^-$.  Therefore, we can plot $G(E)$ as the sum of a sequence of bivalued functions of $E$; see Figure \ref{fig:SingleUnitDots_Comb}(B) and~\cite{Goldman:2003ge}.  The contribution from each pool is then represented by the shaded region in the plot.  Finally, the bias terms are chosen to center these shaded boxes over the identity line. The correct biases in this general case are:
\begin{equation}
	b_i = p\left(\frac{r^-+r^+}{2}\right) + q((i-1)r^+ + (N-i)r^-) \;.
\end{equation}

\subsection{Fixation lines}

%FIG*FIG*FIG*FIG*FIG*FIG*FIG*FIG*FIG*FIG*FIG*FIG*FIG*FIG*FIG*FIG*FIG*FIG*FIG*FIG
\begin{figure}
	\makebox[6in][l]{(A)\hspace{3in}(B)}
	\begin{minipage}[b]{3in}
		\includegraphics[width=3in]{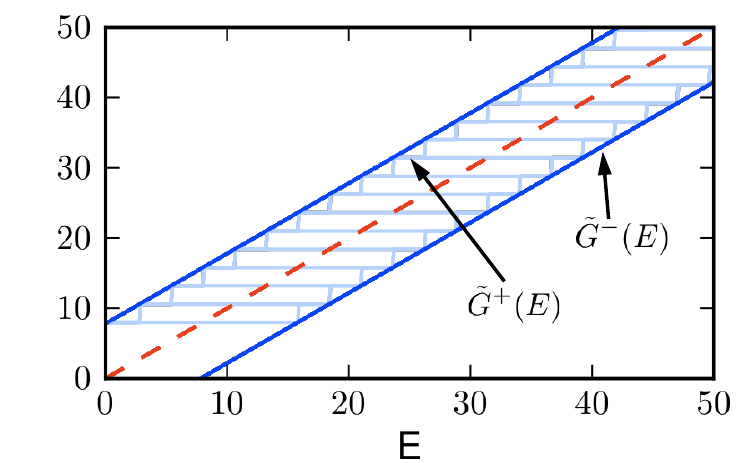}
	\end{minipage}
	\begin{minipage}[b]{3in}
		\centering
		\includegraphics[width=3in]{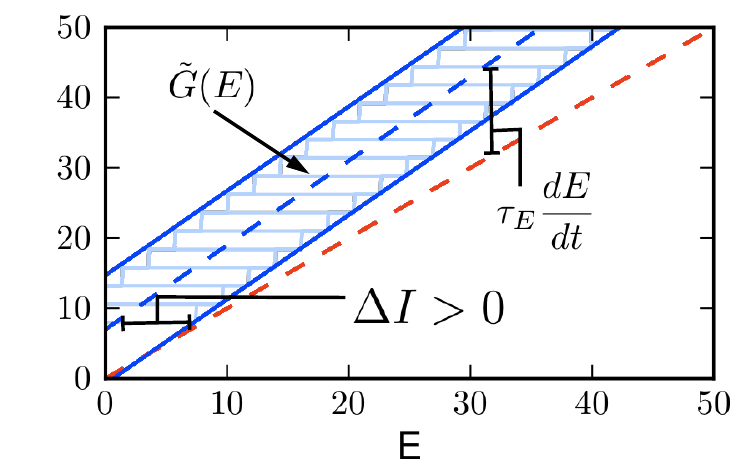}
	\end{minipage}
	\caption{Plot of possible equilibria for Equation \ref{eq:sumBothSides}. (A) The extent of each multivalued feedback function defines the minimum input necessary to perturb the system away from equilibrium, defining the ``fixation" lines.  As $N\rightarrow\infty$, the stable fixed points become more tightly packed on the interval $(r^-,r^+)$.  (B) When the integrator is mistuned, the fixation lines and the feedback line are no longer parallel.  The rate that the integrator accumulates input is approximated by the distance between the center line of the feedback subunits ($\tilde{G}(E)$), and the feedback line.  However, integration only occurs when the ``fixation condition" is no longer satisfied, i.e. when the feedback line is no longer bounded by the fixation lines. }
	\label{fig:fixation}
\end{figure}
%FIG*FIG*FIG*FIG*FIG*FIG*FIG*FIG*FIG*FIG*FIG*FIG*FIG*FIG*FIG*FIG*FIG*FIG*FIG*FIG

We next define the ``fixation" lines ($\tilde{G}^+(E)$ and $\tilde{G}^-(E)$), which are the consequence of this multi-valued property of the integrator. These lines define a region (the fixation region) that runs across the outermost corners of the ``stacked boxes" in Figure \ref{fig:fixation}.  
\begin{equation}
	\tilde{G}^{+} (E) = E(1+\beta) + \frac{a\Delta I}{Nq}-\frac{\beta r^+}{N} + \frac{p(r^+-r^-)}{2Nq}
\end{equation}
\begin{equation}
	\tilde{G}^{-} (E) = E(1+\beta)+ \frac{a\Delta I}{Nq} -\frac{\beta r^-}{N} - \frac{p(r^+-r^-)}{2Nq} \;.
\end{equation}
The term fixation region refers to the following property:  if the input $\Delta I$ is such that the identity line lies within the fixation region, then the integrator will possess a range of closely spaced fixed points (where $E=G(E)$).   Thus, $E$ is not expected to change from its current value; in other words, integration of $\Delta I$ will not occur~\cite{Goldman:2003ge,Koulakov:2002kx}.  Recall that $\Delta I$ acts to shift these boxes leftward or rightward relative to the identity line, just as in the analysis of Figure \ref{fig:SingleUnitDots_Comb}.  As a consequence, it is weak inputs that fail to be integrated.

From this analysis, we can see that integration by the system as a whole relies on two concepts.  The first is a condition on $\Delta I$ necessary to eliminate fixed points; we call this the ``fixation condition."  The second is the question of how quickly to integrate once this condition is no longer satisfied.  We address this next.

\subsection{Integration}

Based on the analysis above, we derive a reduced model that approximately captures the dynamics of the ``full" model indicated by Equation \ref{eq:sumBothSides}.  We call this the ``effective" model.  The rate of change of $E$ -- i.e., the rate of integration -- is given by the distance between the the current value of $G(E)$ and the identity line.  We approximate  this by the distance between the {\it middle} of the fixation lines, which we define as $\tilde G(E)$, and the identity line.  This is pictured in Figure \ref{fig:fixation}(B), and yields:

\begin{eqnarray}
	\tau_E \frac{dE}{dt} &=& -E + G(E) \approx -E + \tilde{G}(E) \\
	\tilde{G}(E) &=& (1+\beta)E + \frac{a\Delta I}{Nq} - \beta\frac{(r^++r^-)}{2N}
\end{eqnarray}
We emphasize that integration by this equation only occurs when the ``fixation" condition is no longer satisfied, i.e. when the fixation lines no longer bound the the identity line.
	
\subsection{Fixation condition}

The last step in defining the 1-dimensional ``effective" model is determining the fixation condition. We must solve for the values of $E$ that cause the feedback line to lie between the two fixation lines:
%	the formulas for these lines are simply vertically shifted versions of $\tilde{G}$ lines.  Requiring both of these conditions be satisfied is equivalent to requiring $E$ be such that:

\begin{eqnarray}
	\text{No Change in E} &\iff& \tilde{G}^-(E) < E < \tilde{G}^+(E)\\
	 &\iff& \tilde{G}(E) - \frac{(r^+-r^-)(p-q\beta)}{2Nq} < E < \tilde{G}(E) + \frac{(r^+-r^-)(p-q\beta)}{2Nq}\\
	&\iff& \left| \beta E + \frac{a\Delta I}{Nq} - \beta \frac{(r^++r^-)}{2N} \right| < \frac{(r^+-r^-)(p-\beta q)}{2Nq}
\end{eqnarray}
If this condition is violated with $\Delta I=0$, the integrator displays runaway integration ($\beta>0$) or leak ($\beta<0$).  If it is satisfied when $\Delta I=0$, we have a condition on the level of $\Delta I$ that must be present for integration to occur.  This yields a piecewise-defined differential equation, corresponding to when integration can and cannot occur:
	
	%EQN*EQN*EQN*EQN
 \begin{displaymath}
   \tau_E\frac{dE}{dt} \approx \left\{
     \begin{array}{ll}
       0  & :  \left| \beta E + \frac{a\Delta I}{Nq} - \beta\frac{(r^++r^-)}{2N} \right| < \frac{(r^+-r^-)(p-\beta q)}{2Nq} \\
       \beta E  + \frac{a \Delta I}{Nq} - \beta\frac{(r^++r^-)}{2N} & : \;\;\;\;\;\;\;\;\;\;\;\; \text{otherwise}
     \end{array}
   \right.
   \label{eq:effective}
\end{displaymath}
	%EQN*EQN*EQN*EQN
	
We now simplify this equation as follows.  First, we assume that $p$ is  much larger than $\beta q$, so that $\frac{(r^{+}-r^{-})(p-\beta q)}{2Nq} \approx \frac{(r^{+}-r^{-})p}{2Nq}  = R$, where $R$ is the robustness parameter in the main text.  Next, we note that as  $N$ increases, the additive term $\beta\frac{(r^++r^-)}{2N}$ becomes negligible (here the synaptic weights $a$ and $p$ can adjusted so that the remaining terms do not vanish).  Finally, we define $\kappa = \frac{a}{N q}$.  This yields the central Equation 5 of the main text:
 \begin{equation}
 \boxed{
   \tau_E \frac{dE}{dt} = \left\{
     \begin{array}{ll}
       0  & :  \left| \beta E + \kappa \Delta I \right| \leq R \\
        \beta E + \kappa I & :  \text{otherwise}
     \end{array}
   \right.}
      \label{eq:chopModelZeroBeta}
\end{equation}
%This equation provides the starting point for the mathematical analysis in the main text.

% - - - - - - - - - - - - - - - - - - - - - - - - - - - - - - - - - - - - - -
\subsection{Quality of the model reduction}

%FIG*FIG*FIG*FIG*FIG*FIG*FIG*FIG*FIG*FIG*FIG*FIG*FIG*FIG*FIG*FIG*FIG*FIG*FIG*FIG
\begin{figure}
	\includegraphics[width=6in]{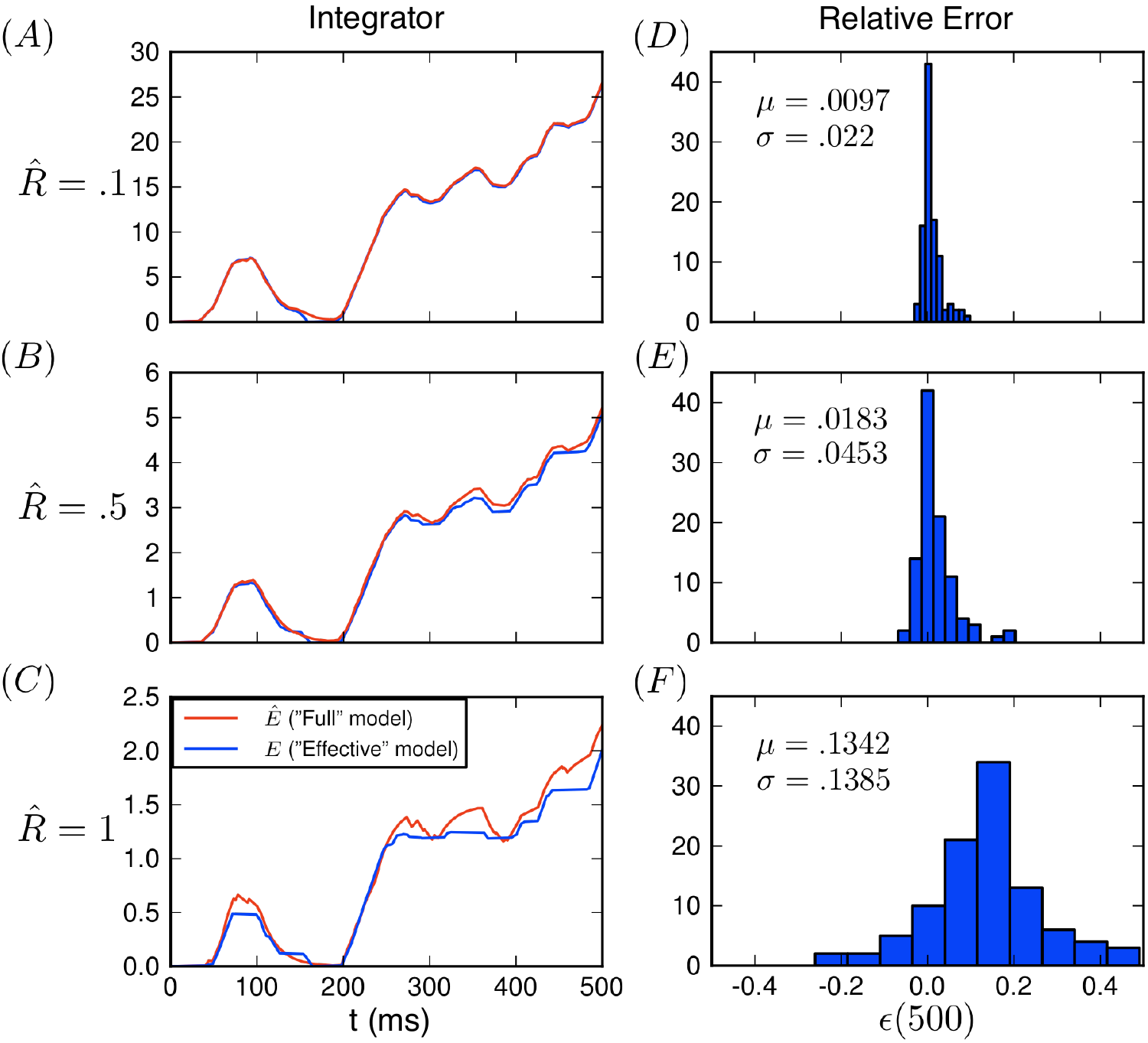}
	\caption{Quality of the reduction from the ``full" model to the ``effective" model. For three different values of the robustness limit $\hat{R}$, traces $E(t)$ for the full model (Equation \ref{eq:fullSystem}), and its reduction $\hat E(t)$ (Equation \ref{eq:effective}) are compared (Panels A, B, and C).  Here, the same signal is integrated by each model.  Panels D, E, and F show histograms (N=100) of the relative error between the two models at $t=500$ ms. (See Equation \ref{eq:relativeError}). The mean $\mu$ and standard deviation $\sigma$ for each distribution are also indicated. At these levels of $\hat{R}$, the effective model gives an approximation of the full subpopulation-based model with low to moderate error (see text). }
	\label{fig:3Traces3Hist}
\end{figure}
%FIG*FIG*FIG*FIG*FIG*FIG*FIG*FIG*FIG*FIG*FIG*FIG*FIG*FIG*FIG*FIG*FIG*FIG*FIG*FIG

Figure \ref{fig:3Traces3Hist} compares the reduced ``effective" model, defined by Equation \ref{eq:chopModelZeroBeta} and denoted by the variable $E$, and the ``full model" of multiple subunits, defined by Equation \ref{eq:fullSystem} and denoted by the variable $\hat{E}$. Results are given for three different values of the normalized robustness limit $\hat{R}$ \footnote{This is the robustness parameter divided by the steady-state standard deviation of the input signal; see main text.} are displayed, all in response to the same input signal $\Delta I(t)$ for ease of comparison.  As $\hat{R}$ increases, the ability of the effective model to track the full model decreases.  The quality of the reduction can be quantified by examining the relative error between the full and effective models:
\begin{equation}
	\epsilon(t) = \frac{\hat{E}(t)-E(t)}{\hat{E}(t)}
	\label{eq:relativeError}
\end{equation}
Histograms of $\epsilon(t)$ evaluated at $t=500$ sec. are given in Figure \ref{fig:3Traces3Hist}(D-F).  The agreement of the two models is within roughly 20\% across a range of robustness values $\hat R$.  This is sufficient for our purposes of demonstrating an approximate connection between the simplified integrator model used in the main text and one of its many possible neural substrates.

\section{Additional indicators that robust integration is compatible with empirical data from decision tasks}
\label{sec:appendix3}

As emphasized in the main text, we make several assumptions about how evidence is integrated over time in our model of perceptual decision making.  The first is that the tuning of connectivity in the integrator circuit is inherently imprecise from trial to trial.  This imprecision in feedback tuning results in spurious activity, in the form of either exponential growth or decay.  To counter this, we consider the presence of a robustness mechanism meant to ameliorate this problem by allowing the integrator to fixate at various levels of activity despite mistuning.

These assumptions differ from, e.g., the drift-diffusion model for two-alternative decision making, which has been shown to explain psychophysical and physiological data.  An important question is therefore whether the model at hand will also be able to capture these data.  In the main text we find that the answer is yes, for behavioral data on decision accuracy and speed (Figures 14 and 16 in the main text).  We next expand this analysis by considering several other aspects of the data:  error-trial vs. correct-trial reaction time histograms, controlled duration accuracy performance, decision-triggered stimulus predictions, and traces of neural firing rates vs. time.

\subsection{Asymmetry of reaction times between correct and incorrect trials}

%FIG*FIG*FIG*FIG*FIG*FIG*FIG*FIG*FIG*FIG*FIG*FIG*FIG*FIG*FIG*FIG*FIG*FIG*FIG*FIG
\begin{figure}
	\makebox[6in][l]{(A)\hspace{3in}(B)}
	\begin{minipage}[b]{3in}
		\centering
		\includegraphics[width=3in]{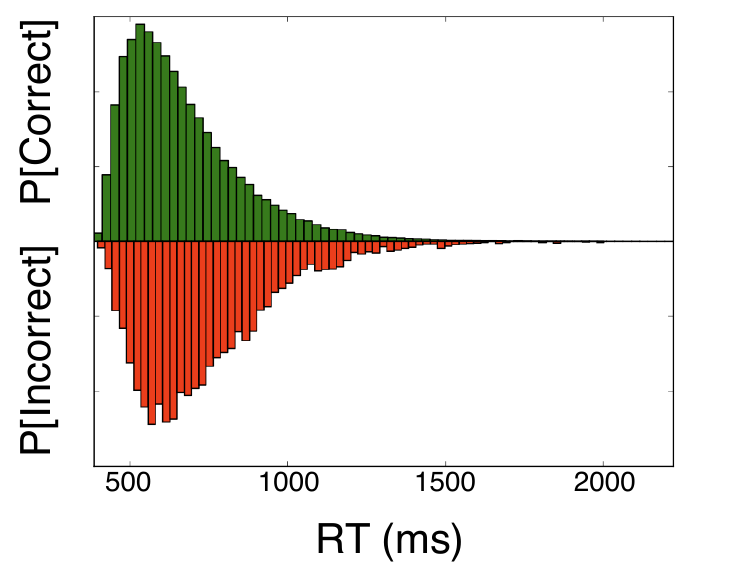}
	\end{minipage}
	\begin{minipage}[b]{3in}
		\includegraphics[width=3in]{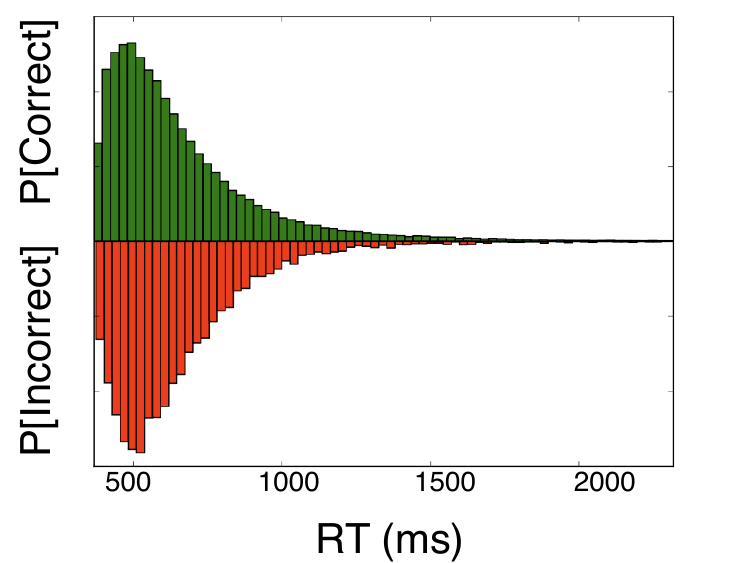}
	\end{minipage}
	\caption{Reaction time histograms from Figures 15 and 16 of the main text, sorted by correct and incorrect responses and plotted in Panels (A) and (B), respectively. Correct trials are pictured above, while incorrect trials are below; settings as in the table included in Figure 14 of the main text. (A) Mean reaction time for error trials is 740 ms., vs. 671 ms. for correct trials; here a non-robust integrator is precisely tuned ($\sigma_{\beta}=0$, and $\hat{R}=0$) (B) For a robust integrator with imprecise tuning, error trials are now shorter (644 vs. 652 ms.).}
	\label{fig:RTAsymmetry}
\end{figure}
%FIG*FIG*FIG*FIG*FIG*FIG*FIG*FIG*FIG*FIG*FIG*FIG*FIG*FIG*FIG*FIG*FIG*FIG*FIG*FIG

\citetext{Roitman:2002wr} and many other authors report differences in mean reaction times when responses are sorted by correct vs. incorrect responses.  In Figure~\ref{fig:RTAsymmetry}, we perform the same analysis on the data in Figure~14 of the main text. First, Panel (A) shows that our model of a perfectly tuned, non-robust integrator produces the shows the same qualitative asymmetry, with error trials on average having longer reaction times.  With the addition of imprecise tuning and robustness, however this asymmetry can be eliminated (Panel (B)).

Numerous changes to the model could recover the longer error RTs.  One possibility is to introduce a bias in mistuning parameters:  choosing $\bar{\beta}>0$ has the desired effect.  Others include inclusion of collapsing bounds, urgency signals, and other forms of trial-to-trial variability (for example \citetext{Ratcliff:1998vl}).

%which .  When $\bar{\beta}<0$, this inconsistent asymmetry is exacerbated, and when $\bar{\beta}>0$, it reverts back to the case consistent with observations (Both results not shown).  We conclude that this analysis suggests that the model without assumptions is consistent with reported results. However, when robustness and imprecise tuning are assumed, the imprecision in tuning must be biased positive to maintain this consistency.

\subsection{Model consistency with controlled duration data}
\label{sec:KianiMatch}

In the main text, we showed that an integrator model with imprecise tuning and robustness can reproduce the chronometric and psychometric curves reported in \citetext{Roitman:2002wr} for the free-response task.  We also considered the consequences of both robustness and mistuning in a controlled duration task.
Here we show that the {\it same} parameters (Table, Figure 14 of the main text), this model can reproduce behavioral data reported in \citetext{Kiani:2008ee} for the controlled duration task.  

\begin{figure}
	\makebox[6in][l]{(A)\hspace{3in}(B)}
	\begin{minipage}[b]{3in}
		\centering
		\includegraphics[width=3in]{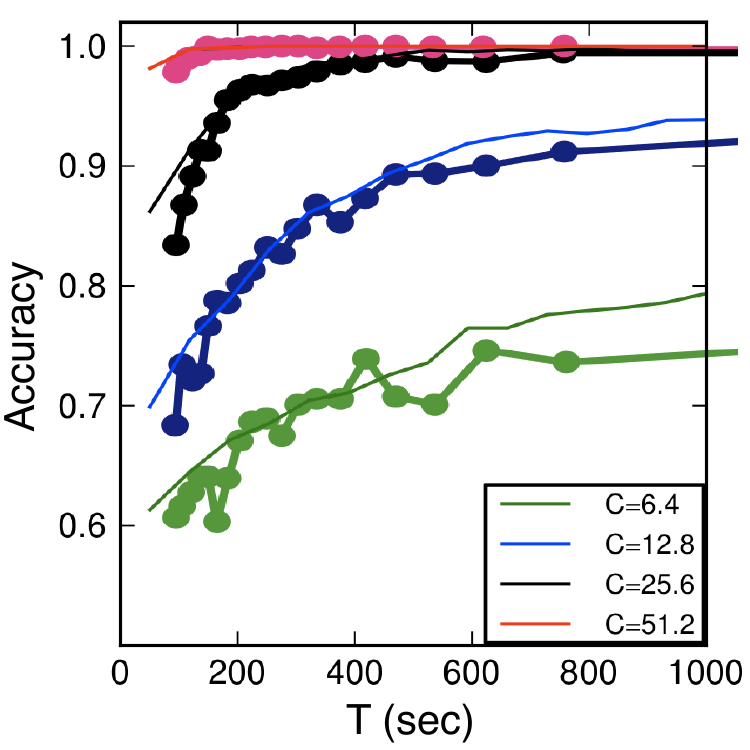}
	\end{minipage}
	\begin{minipage}[b]{3in}
		\includegraphics[width=3in]{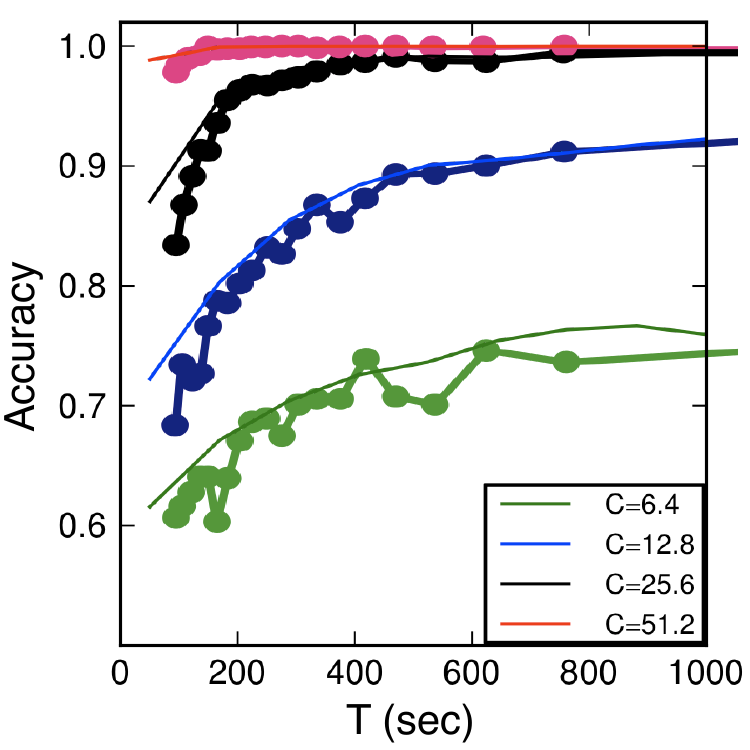}
	\end{minipage}
	\caption{Comparison of model with controlled duration accuracy, C=12.8.  The perfectly tuned model (A), and imprecise/robust model (B), both predict accuracy that is comparable with those reported in \protect\cite{Kiani:2008ee} (Dotted lines, overlaid).  In both cases, the CD task is modeled with bounded integration, with threshold set as in the table of Figure 14, main text.}
	\label{fig:KianiCompare}
\end{figure}

We assume bounded integration, where an integration threshold determines which alternative is selected, and the subject waits until the end of the trial to report the selection.  The comparison is demonstrated in Figure~\ref{fig:KianiCompare}, where model the accuracy functions from the model are overlaid on the data provided in the Kiani et al. study, with no parameter changes (including threshold) from the data reported in the fit to the Roitman et al. reaction time task data in the main text.

%FIG*FIG*FIG*FIG*FIG*FIG*FIG*FIG*FIG*FIG*FIG*FIG*FIG*FIG*FIG*FIG*FIG*FIG*FIG*FIG
%FIG*FIG*FIG*FIG*FIG*FIG*FIG*FIG*FIG*FIG*FIG*FIG*FIG*FIG*FIG*FIG*FIG*FIG*FIG*FIG

%FIG*FIG*FIG*FIG*FIG*FIG*FIG*FIG*FIG*FIG*FIG*FIG*FIG*FIG*FIG*FIG*FIG*FIG*FIG*FIG
\begin{figure}
	\makebox[6in][l]{(A)\hspace{3in}(B)}
	\begin{minipage}[b]{3in}
		\centering
		\includegraphics[width=3in]{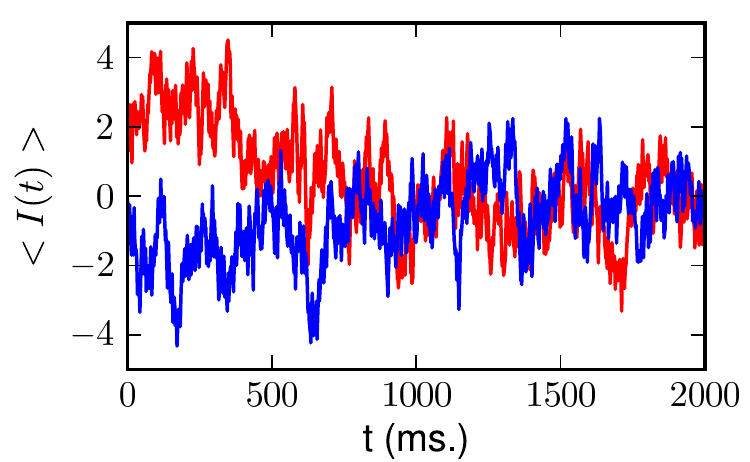}
	\end{minipage}
	\begin{minipage}[b]{3in}
		\includegraphics[width=3in]{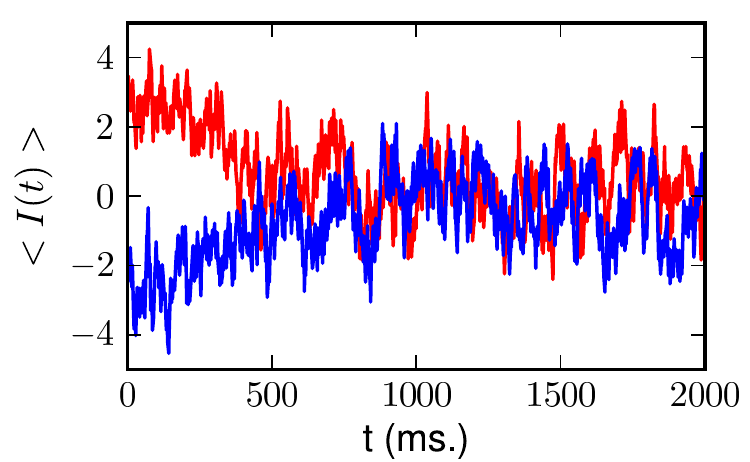}
	\end{minipage}
	\caption{Time course of decision-triggered stimuli. Stimulus realizations $\Delta I(t)$ are sorted by ``left'' vs ``right" alternative selected (red vs. blue; neither alternative is ``correct" as coherence C=0), and averaged across trials.  Again assuming perfect tuning, (A), and imprecise tuning with robustness, (B), we find that both results are consistent with the motion energy profiles reported in \protect\citetext{Kiani:2008ee}. All integrator settings are the same as reported in the table in Figure 14 of the main text.}
	\label{fig:DTSP}
\end{figure}
%FIG*FIG*FIG*FIG*FIG*FIG*FIG*FIG*FIG*FIG*FIG*FIG*FIG*FIG*FIG*FIG*FIG*FIG*FIG*FIG

\subsection{Decision-triggered stimuli}
\citetext{Kiani:2008ee} also report the time-course of motion evidence leading to the eventual decision in the controlled duration task.  Specifically, the authors study long duration trials with neutral stimuli (i.e., zero coherence, so neither alternative is ``correct").  In our model, motion evidence is encoded by the stimulus $\Delta I(t)$ (see main text).  In Figure \ref{fig:DTSP} we sort 400 trials (coherence C=0) based on which alternative was selected.  We average the stimuli, again assuming bounded integration in a controlled duration task.  

We observe that the results of our model, both perfectly tuned and imprecisely tuned with nonzero robustness, are consistent with those reported by Kiani et al. Specifically, the early separation of motion energy profiles for rightward (red) and leftward (blue) choices are indicative bounded accumulator, and do not qualitatively change when mistuning and robustness are included in the model.

\subsection{Trial-sorted traces of firing rates vs. time}

%FIG*FIG*FIG*FIG*FIG*FIG*FIG*FIG*FIG*FIG*FIG*FIG*FIG*FIG*FIG*FIG*FIG*FIG*FIG*FIG
\begin{figure}
	\makebox[6in][l]{(A)\hspace{3in}(B)}
	\begin{minipage}[b]{3in}
		\centering
		\includegraphics[width=3in]{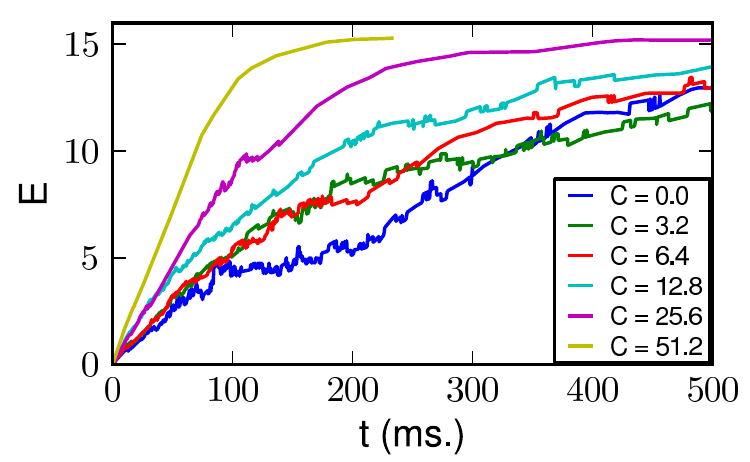}
	\end{minipage}
	\begin{minipage}[b]{3in}
		\includegraphics[width=3in]{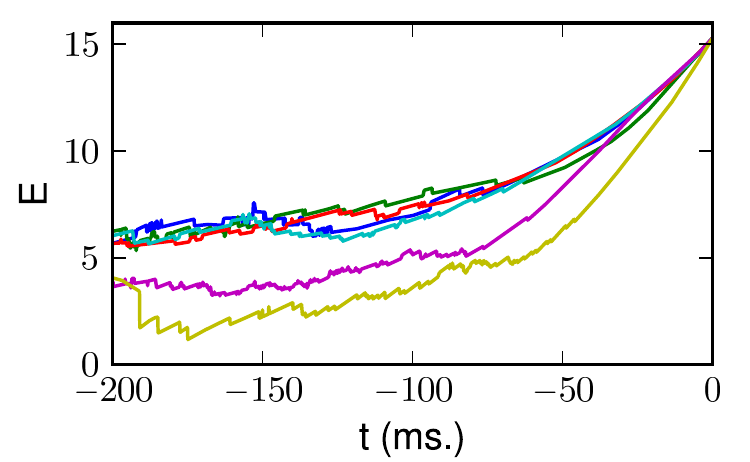}
	\end{minipage}
	\makebox[6in][l]{(C)\hspace{3in}(D)}
	\begin{minipage}[b]{3in}
		\centering
		\includegraphics[width=3in]{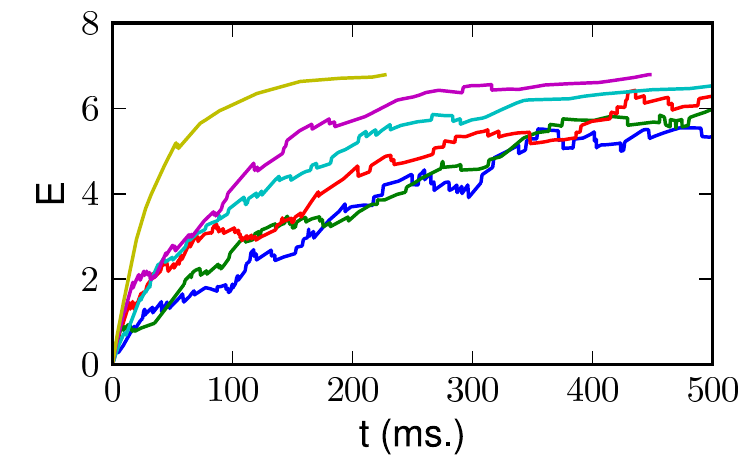}
	\end{minipage}
	\begin{minipage}[b]{3in}
		\includegraphics[width=3in]{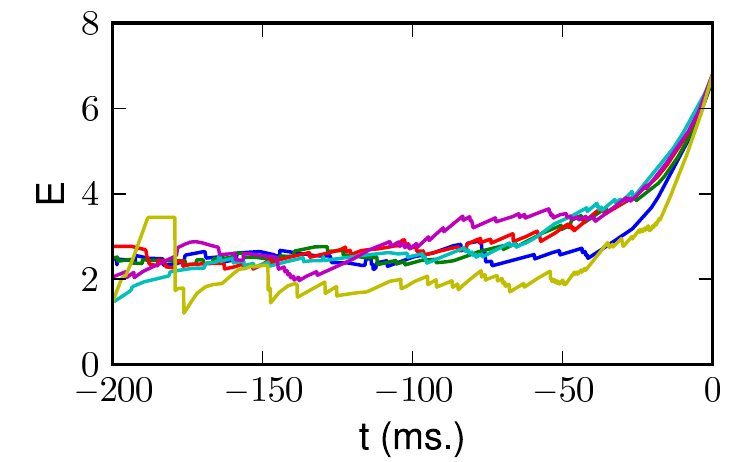}
	\end{minipage}
	\label{fig:TAIT}
	\caption{Time course of model integrator activity $E(t)$, during stimulus integration in reaction time tasks.  Integration of input signal at various coherence values, with responses aligned at the beginning (A, C) and end (B, D), averaged across 400 trials.  As expected, activity ramps up faster as dot coherence increases in both the perfectly tuned (A, B) and imprecisely tuned/robust (C, D) cases, in accordance with physiological observations. (For example, \protect\citetext{Roitman:2002wr}.)}
\end{figure}
%FIG*FIG*FIG*FIG*FIG*FIG*FIG*FIG*FIG*FIG*FIG*FIG*FIG*FIG*FIG*FIG*FIG*FIG*FIG*FIG

Firing rates of LIP neurons reflect the time course of accumulated sensory evidence~\cite{Gold:2007fo}.  We verified that our model displays analogous ramping activity when averaged over the similar number of trials.  In particular, we checked this both for the perfectly tuned, non-robust version of our model (Figure 7(A-B)) and in the presence of mistuning and robustness (Figure 7(C-D)).

Integrator activity averaged across multiple trials is averaged, and aligned to both the onset of stimulus and upon reaching threshold in both model setups. Activity ramps up (integrates) faster at higher dot coherence values, consistent with physiological measurements. This provides further evidence that our two additional model assumptions are consistent with known physiology for circuits involved in perceptual decision making.

\bibliographystyle{jneurosci}
\bibliography{robustIntegratorSupplement}